\documentclass[letterpaper,twocolumn,10pt]{article}

\usepackage[available,functional,reproduced]{usenixbadges}
\usepackage{usenix-2020-09}
\usepackage{booktabs} 
\usepackage{graphicx} 

\usepackage[normalem]{ulem}
\usepackage{authblk}

\usepackage{caption}
\usepackage{subcaption}
\usepackage{balance} 
\usepackage{array}
\usepackage{makecell,multirow}

\usepackage{marvosym} 

\usepackage{xspace}
\newcommand{\untrustedos}{{{primary OS}}\xspace}
\newcommand{\visorname}{{{RustMonitor}}\xspace}
\newcommand{\host}{{{monitor mode}}\xspace}
\newcommand{\normal}{{{normal mode}}\xspace}
\newcommand{\secure}{{{secure mode}}\xspace}

\newcommand\yuekai[1]{\textcolor{cyan}{\{\textbf{yuekai:} {\em#1}\}}}

\newcommand\shoumeng[1]{\textcolor{blue}{\{\textbf{shoumeng:} {\em#1}\}}}

\newcommand\wenhao[1]{}
\newcommand\shuang[1]{}

\newcommand{\mypara}[1]{\vspace{0pt}\noindent\textbf{{#1.}}}

\newcommand{\para}[1]{\vspace{0pt}\noindent\textbf{{#1.}}}
\newcommand{\ignore}[1]{}  

\newcommand{\atc}[1]{{{#1}}}

\newcommand{\PreserveBackslash}[1]{\let\temp=\\#1\let\\=\temp}
\newcommand{\circled}[1]{\raisebox{.5pt}{\textcircled{\raisebox{-.9pt} {#1}}}}

\newcolumntype{C}[1]{>{\PreserveBackslash\centering}p{#1}}
\newcolumntype{R}[1]{>{\PreserveBackslash\raggedleft}p{#1}}
\newcolumntype{L}[1]{>{\PreserveBackslash\raggedright}p{#1}}

\newcounter{packednmbr}

\newenvironment{packeditemize}{
\begin{list}{$\bullet$}{
\setlength{\labelwidth}{3pt} 
\setlength{\itemsep}{0pt}
\setlength{\leftmargin}{\labelwidth}
\addtolength{\leftmargin}{\labelsep}
\setlength{\parindent}{0pt}
\setlength{\listparindent}{\parindent}
\setlength{\parsep}{2pt} 
\setlength{\topsep}{2pt}}}{\end{list}}

\title{\Large \bf HyperEnclave: An Open and Cross-platform Trusted Execution Environment }
\begin{document}

\author[1]{\rm Yuekai Jia}
\author[2]{\rm Shuang Liu}
\author[3,4(\Letter\thanks{Corresponding author: Wenhao Wang (\href{mailto:wangwenhao@iie.ac.cn}{wangwenhao@iie.ac.cn}).})]{\rm Wenhao Wang}
\author[1]{\rm Yu Chen}
\author[2]{\rm Zhengde Zhai} 
\author[2]{\\ \rm Shoumeng Yan}
\author[2]{\rm Zhengyu He}
\affil[1]{\it Tsinghua University}
\affil[2]{\it Ant Group}
\affil[3]{\it SKLOIS, Institute of Information Engineering, CAS}
\affil[4]{\it School of Cyber Security, University of Chinese Academy of Sciences}

\pagestyle{empty} 
\maketitle


\begin{abstract}
A number of trusted execution environments (TEEs) have been proposed by both academia and industry. However, most of them require specific hardware or firmware changes and are bound to specific hardware vendors (such as Intel, AMD, ARM, and IBM). 
In this paper, we propose HyperEnclave, an open and cross-platform process-based TEE \shuang{todo: why process-based ? need more justifications}that relies on the widely-available virtualization extension to create the isolated execution environment. 
In particular, HyperEnclave is designed to support the \textit{flexible enclave operation modes} \shuang{todo: what's the key benefits of p-enclave compared with vm-based TEE ? real work load to support P-enclave;  }to fulfill the security and performance demands under various enclave workloads. 
We provide the enclave SDK to run existing SGX programs on HyperEnclave with little or no source code changes.
We have implemented HyperEnclave on commodity AMD servers and deployed the system in a world-leading FinTech company to support real-world privacy-preserving computations.
The evaluation on both micro-benchmarks and application benchmarks shows the design of HyperEnclave introduces only a small overhead.

\end{abstract}


\section{Introduction}
\label{sec:intro}

In recent years, trusted execution environments (TEEs) are emerging as a new form of computing paradigm, known as confidential computing, due to the high demand for privacy-preserving data processing technologies that can handle massive data samples. TEEs provide hardware-enforced memory partitions where sensitive data can be securely processed. Existing TEE designs support different levels of TEE abstractions, such as process-based (Intel's Software Guard eXtensions (SGX)~\cite{mckeen2013innovative}), VM-based (AMD SEV~\cite{kaplan2016amd}), separate worlds (ARM TrustZone~\cite{2004Trustzone}), and hybrid (Keystone~\cite{lee2020keystone}). 
Currently, the most prominent example of TEEs is Intel SGX, which is widely available in commercial off-the-shelf (COTS) desktop and server processors. 

\mypara{Motivations}
Most of today's TEE technologies are close-sourced and require specific hardware or firmware changes that are difficult to audit, slow to evolve, and thus are inferior to cryptographic alternatives (such as homomorphic encryption), which are based upon public algorithms and widely available hardware. 
Moreover, most existing TEE designs restrict the enclaves (i.e., the protected TEE regions) to run only in fixed mode.\footnote{An exception is CURE~\cite{bahmani2021cure}, which however requires hardware changes to the CPU core and the system bus to support the flexible enclaves (Sec.~\ref{sec:related}).} 
\ignore{
keystone:
We showcase how Keystone-based TEEs run on unmodified RISC-V hardware...

CURE:
The Rocket core supports three software privileged level (user, supervisor and machine). 
Our CURE prototype implements user-space enclaves, kernel-space enclave and sub-space enclaves ... 
the enclave-to-peripheral binding requests hardware change. 
}
It is difficult to support the performance and security requirements of various types of applications that need to be protected by TEEs. 
For example, Intel SGX enclaves run in the user mode and cannot access privileged resources (such as the file system, the IDT, and page tables) and process privileged events (interrupt and exceptions). 
As a result, running I/O-intensive and memory-demanding tasks leads to significant performance degradation. \shuang{todo: consolidate scenarios }
\ignore{
how related work promote flexibility ? 
CURE: 
existing TEE solutions suffer from significant design shortcomings. First, they follow a one-size-fit-all approach offering only a single enclave type., however, different services need flexible enclaves that can adjust to their demands.  Second, they cannot efficiently support emerging applications (such as Machine-learn as a service) , which requires secure channels to peripherals (e.g. accelerator). 

Keystone: 
Each vendor TEE enables only a small portion of the possible design space across threat models, hardware requirements, resource management, porting effort, and feature compatibility. 
 when a enclave chooses a target hardware platform, they are locked into the respective TEE design limitations regardless of their actual application needs. 
we advocate that the HW must provide security primitives and not point-wise solutions. 
}

To fill the gap,
in this paper we propose the design of HyperEnclave to support \textit{confidential cloud computing} that can run securely on both legacy servers readily available in the cloud, and on the rising ARM (or RISC-V in the future) servers, without requiring specific hardware features. For this purpose, our design provides a process-based TEE abstraction using the widely available virtualization extension (for isolation) and TPM (for root of trust and randomness etc.). 
To better fulfill the needs for specific enclave workloads, HyperEnclave supports the \textit{flexible enclave operation modes}, i.e., the enclaves can run at different privilege levels and can have access to certain privileged resources (see Sec.~\ref{sec:flexibility} for more details).

\mypara{Design details}
In our design, the system runs in three modes. A trusted software layer, called \visorname (security monitor written in Rust),
runs in the \textit{\host}, which is mapped to the VMX root mode. RustMonitor is responsible for enforcing the isolation and is part of the trusted computing base (TCB). The untrusted OS (referred to as the \textit{\untrustedos}) provides an execution environment for the untrusted part of applications; the untrusted OS and application parts run in the \textit{\normal}, which is mapped to the VMX non-root mode. 
The trusted part of application (i.e., \textit{enclave}) runs in the \textit{\secure}, which can be mapped \textit{flexibly} to ring-3 or ring-0 of the VMX non-root mode, or ring-3 of the VMX root mode.

Memory isolation is enforced with hardware-based memory protection of the memory-management unit (MMU). As we observe that existing process-based TEEs (e.g., Inktag~\cite{Hofmann2013InkTagSA} and Intel SGX~\cite{mckeen2013innovative}) are vulnerable to page-table-based attacks~\cite{xu2015controlled}, our memory isolation scheme chooses to manage the enclave's page table and page fault events entirely by the trusted code, removing the involvement of the \untrustedos. The design also prevents certain types of enclave malware attacks (Sec.~\ref{sec:enclave-memory-protection}). 


To minimize the attack surface, we adopt an approach called \textit{measured late launch}: the \untrustedos kernel is first booted; then a chunk of special kernel code, implemented as a kernel module in the \untrustedos , runs to initiate \visorname in the most privileged level (i.e., the monitor mode) and demotes the \untrustedos to the normal mode. All booted components during the booting process are measured and extended to the TPM Platform Configuration Registers (PCRs).
Since the TPM attestation guarantees that PCRs cannot be rolled back, the design ensures that \visorname is securely launched; 
otherwise, a violation of the TPM \textit{quote} would be detected during remote attestation.

We have implemented HyperEnclave on commodity AMD servers. In total \visorname consists of about 7,500 lines of Rust code. 
The APIs of our enclave SDK are compatible with the official SGX SDK.
As a result, code written for SGX could be easily ported to run on HyperEnclave by recompiling the code with little (or no) source code changes. 
We have ported a number of SGX applications, as well as the Rust SGX SDK~\cite{Wang2019TowardsMS} and the Occlum library OS~\cite{shen2020occlum} to HyperEnclave.
The micro-benchmarks show that the overheads for ECALLs and OCALLs are < 9,700 and < 5,260 cycles respectively (14,432 and 12,432 cycles respectively on Intel SGX).
The evaluation on a suite of 
real-world applications shows that the overhead is small (e.g., the overhead on SQLite is only 5\%). 

\mypara{Contributions}
In summary, the paper proposes the design of HyperEnclave, with the following contributions:
\begin{packeditemize}
    \item An open\footnote{The code will be available at \href{https://github.com/HyperEnclave}{https://github.com/HyperEnclave}.
    } and cross-platform processed-based TEE with minimum hardware requirements (virtualization extensions and TPM) that can run existing SGX programs with little or no source code changes\atc{, which enables the reuse of the rich toolchains and ecosystem for Intel SGX}. 
    \item Supporting the flexible enclave operation modes to fulfill the diverse security and performance requirements of enclave applications without hardware or firmware changes. 
    \item A memory isolation scheme that the enclave's page table and page fault are managed entirely by the trust code, which mitigates the page-table-based attacks and the enclave malware attacks. 
    \item A measured late launch approach, combined with the TPM-based attestation to reduce the attack surface.
    \item An implementation on commodity servers (mostly) using the memory safe language Rust, and an evaluation on real hardware and applications, demonstrating that the proposed design is practical and only has a small overhead.
\end{packeditemize}


\section{Background}

\subsection{Trusted Execution Environment}

A Trusted Execution Environment (TEE) is designed to ensure that sensitive data is stored, processed, and protected in an isolated and trusted environment. The isolated area could be a separate system apart from the normal operating system (such as the TrustZone~\cite{2004Trustzone} secure world), a part of a process address space (such as an Intel SGX~\cite{mckeen2013innovative} enclave), or a stand-alone VM (such as a virtual machine protected by AMD SEV~\cite{kaplan2016amd} or Intel TDX~\cite{tdx2020}).
To resist the privileged attacker, TEE needs to thwart not only the OS-level adversary but also the malicious party who has physical access to the platform. To this end, it offers hardware-enforced security features including isolated execution, integrity, and confidentiality protection of the enclave, along with the ability to authenticate the code running inside a trusted platform through remote attestation.

\mypara{Isolation}
At the core of a TEE is the memory isolation scheme, which guarantees that code, data, and the runtime state of the enclave cannot be accessed or tampered with by untrusted parties. For Intel SGX, the protected memory (i.e., the \textit{enclave}) is mapped to a special physical memory area called Enclave Page Cache (EPC), which is encrypted 
and cannot be directly accessed by other software, firmware, BIOS, and direct memory access (DMA).



\mypara{Attestation}
The goal of remote attestation is to generate an attestation \textit{quote}, which includes the measurement of the software state, signed with the attestation key embedded in the hardware. The remote user verifies the validity of the quote by checking the signature (which reflects the hardware identity) and the measurement (which proves the software state). 


\subsection{Trusted Platform Module}
Trusted Platform Module (TPM) is both an industry-standard~\cite{tpm2016} and an ISO/IEC standard~\cite{tpm2015iso} for a secure cryptoprocessor. 
It is used by nearly all PC and server manufacturers. Firmware TPMs (fTPMs) are firmware-based (e.g. UEFI) TPM implementations. At the time of this writing, Intel, AMD, and Qualcomm all have implemented fTPMs.

TPM has a set of Platform Configuration Registers (PCRs), which can be used for the measurement of the booted code during the boot process. PCRs are reset to zero on system reboot or power on-off. During every boot process, the PCRs can only be extended with the new measurement (called PCR extend), and thus cannot be set to arbitrary values. 

Every TPM ships with a unique asymmetric key, called the Endorsement Key (EK), embedded by the manufacturer as the root of trust. 
The TPM can generate a quote of the PCR values, signed using the TPM Attestation Identity Keys (AIK), while the AIK is generated inside TPM and certified using EK. 
Any modifications of the booted code would be reflected in the quote.
Upon receiving the quote, the remote party can validate the signing key comes from an authentic TPM and can be assured that the PCR digest report has not been altered.

\subsection{Threat Model}
\label{sec:threat-model}
Like the other TEE proposals~\cite{lee2020keystone, 2019SANCTUARY}, we trust the underlying hardware, including the processor establishing the virtualization-based isolation, the System Management Mode (SMM) code, as well as the TPM. We assume that the Core Root of Trust for Measurement (CRTM) is trusted and immutable.
HyperEnclave mitigates certain physical memory attacks, such as cold boot attacks and bus snooping attacks with the hardware support for memory encryption.
\atc{We don't fully trust the operator and assume the attacker cannot mount physical attacks during the boot process, i.e.,} we assume that the system is initially benign (during system boot), and the early OS during the boot stage is part of the TCB. \atc{This can be achieved in two ways.
\begin{packeditemize}
  \item \textit{Firstly}, the power-on event can be secured with a hardware device, such as an HSM (i.e., hardware security module). The platform enters the boot process only with the engagement and supervision of a trusted party, who owns the HSM. After that, the operators for maintenance are not trusted.
  \item \textit{Secondly}, the boot process can be enhanced to defend against adversaries with physical accesses. To prevent I/O attacks,  we can harden the OS to remove unnecessary devices and disable the DMA capability of peripherals before IOMMU is enabled. We can enable memory encryption at an early stage (e.g., in the BIOS, before any off-chip memory is used) to prevent physical memory attacks.
\end{packeditemize}}
However, after \visorname is launched, the \untrustedos is demoted to the normal mode, and can be under the control of the attacker, who may try to compromise the \visorname or enclaves, e.g., try to access the protected memory directly or through DMA. 
We consider the enclave code may be malicious or controlled by an attacker due to memory bugs. Our design needs to prevent a compromised enclave from contaminating the other enclaves or the \visorname. We also prevent the attacks against the \untrustedos or the application code, such as those presented in~\cite{2019practical}.
Similar to other TEEs, in this paper we do not focus on the prevention of denial of service (DoS) attacks or side channel attacks, such as cache timing and speculative execution attacks~\cite{kocher2019spectre}. 
\section{Design}
\label{sec:design}




HyperEnclave is designed to support confidential cloud computing without requiring specific hardware features. Therefore, HyperEnclave is built upon the widely available virtualization extension. 
In particular, HyperEnclave is designed to support the process-based TEE model (similar to Intel SGX) for the following reasons. \shuang{process-based enclave is just one of the enclave types which has pros and cons. Many prior works support it, therefore, I suggest not emphasize it. }
\begin{packeditemize}
    \item \textit{Minimized TCB}. 
    To protect an application using the process-based TEE,
    the TCB includes only the protected code itself, while in the other forms of TEE, much more code must be included, such as the guest operating system for VM-based TEEs. 
    \item \textit{Established ecosystem}. Since Intel SGX is currently the most prevalent TEE supported in the cloud (major CSPs, including GCP, Azure, and Aliyun, provide SGX-based instances~\cite{russinovich2017introducing,asylo2019}), a rich set of toolchains and applications have been developed. Supporting the SGX model reduces the porting effort and makes it easy to deploy confidential computing tasks in the cloud. \shuang{move to SDK ? that's why we support SGX toolchain and the existing SGX apps. }
    \item \textit{Cloud computing trends}. We have witnessed a clear trend towards running container-based serverless applications in the cloud. Protecting these applications against untrusted clouds using TEEs is important. Considering that such computing tasks are typically short-lived, and favor a short start-up time, maintaining a VM seems to be too heavy-weight. 
    \shuang{it's another possible scenario to promote HyperEnclave, refer to PENGLAI for severless computing}
    \ignore{
    microservice[80] and serverless computing[13-15] have become emerging paradigms of cloud, which use single purpose service or function as a basic computation unit and achieve high scalability. 
    }
\end{packeditemize} 
In this section, we introduce HyperEnclave using x86 notations, as we prototyped HyperEnclave on AMD servers. 

\subsection{System Overview}

\begin{figure}[t]
    \centering
    \includegraphics[width=0.90\linewidth]{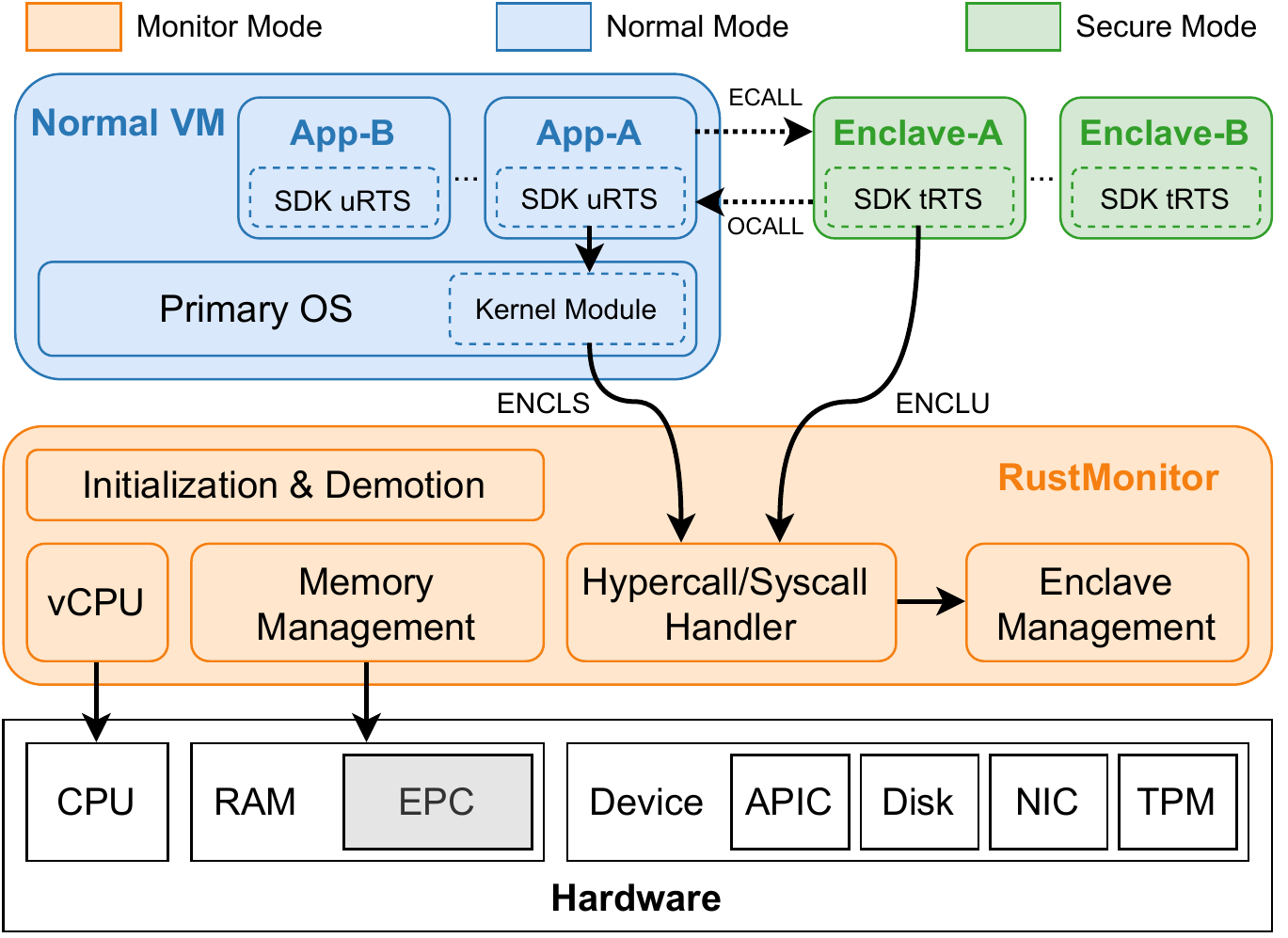}
    \caption{System Overview.}
    \label{fig:overview}
\end{figure}

HyperEnclave supports the following modes: the \host, i.e., VMX root operation mode; the \normal for the \untrustedos and untrusted part of applications, i.e., ring-0 and ring-3 of the VMX non-root operation mode respectively; and the \secure for the enclave, which could be ring-3 and ring-0 of the VMX non-root operation mode, or ring-3 of the VMX root operation mode, depending on the \textit{enclave operation mode}. We will introduce the flexible operation mode supported by HyperEnclave in Sec.~\ref{sec:flexibility}. 
As illustrated in Figure~\ref{fig:overview}, HyperEnclave consists of the following components:




\begin{packeditemize}
    \item \textit{\visorname} is a lightweight hypervisor running in the \host that manages the enclave memory, enforces the memory isolation,
    and controls the enclave state transitions.
    It works as a resource monitor, while complicated tasks are offloaded to the \untrustedos. 
    \item \visorname creates a \textit{unique} guest VM (referred to as the \textit{normal VM}) that runs the \textit{\untrustedos} (such as Linux) and hosts the untrusted part of applications in the \normal. 
    The \untrustedos is still in charge of process scheduling and I/O devices management, but it is not trusted by the \visorname and enclaves. 
    \item \textit{Application} is the untrusted part of the application which runs in the \untrustedos. 
    \item \textit{The kernel module.} We provide a kernel module in the \untrustedos to load, measure, and launch \visorname, as well as to invoke the emulated privileged operations.
    \ignore{
    it loads, measures, and launches Rust-Monitor,
    }
    \item To ease development, HyperEnclave provides an \textit{enclave SDK} with APIs compatible with the official Intel SGX SDK~\cite{sgxsdk}, including both the untrusted runtime and trusted runtime (i.e., SDK uRTS and SDK tRTS). As such, most SGX programs can run on HyperEnclave with little or no source code changes. 
    \item \textit{Enclave} is the trusted part of the application running in the \secure. 
\end{packeditemize}


\subsection{Memory Management and Protection}
\label{sec:enclave-memory-protection}


\noindent\textbf{Challenges.}
%
For process-based TEEs, the enclave runs in the user mode and is not able to manage its own page table. Existing designs (e.g., Intel SGX, TrustVisor~\cite{McCune2010TrustVisorET}) allow the untrusted OS to manage the enclave's page table. To prevent memory mapping attacks (i.e., attacks by manipulating the enclave's address mappings, as shown in Figure~\ref{fig:mapping-attack}, Appendix~\ref{appendix:supplement}), the design of SGX extends the Page Missing Handler (PMH) and introduces a new metadata called EPCM for additional security checks on TLB misses~\cite{Costan2016IntelSE}. 
Without secure hardware support, a prevalent software solution~\cite{Azab2014HypervisionAW,McCune2010TrustVisorET,Yun2019GinsengKS} is to make the page tables write-protected by setting the page table entries (PTEs) for pages holding the page tables, i.e., any update to the page table traps to the hypervisor and then be verified. However, on x86 platforms the updates of access and dirty bits of the PTEs also trap into the hypervisor, leading to non-negligible overhead. Even-worse, since the enclave page fault is also processed by the OS, the above designs are still vulnerable to the page table-based-attacks, such as the controlled-channel attacks~\cite{xu2015controlled}.

The design becomes more challenging to support enclave dynamic memory management (i.e., EDMM on SGX2 platforms~\cite{McKeen2016EDMM}), i.e, dynamically adding or removing enclave pages, or changing the enclave page attributes or types after the enclave is initialized. Without EDMM, all physical memory that the enclave might ever use must be committed before enclave initialization. Therefore, EDMM reduces enclave build time and enables new enclave features, such as on-demand stack and heap growth, and on-demand creation of code pages to support just-in-time (JIT) compilation.
On SGX2 platforms, the enclaves need to send the EDMM request to the SGX driver through OCALLs, who then makes the requested changes. Since the driver is untrusted by the enclaves, the changes need to be explicitly checked and accepted by the enclaves to take effect, which involves heavy enclave mode switches.

\mypara{HyperEnclave memory management}
We observe that the above challenges are rooted in the fact that the enclave's page table and page faults are both managed by the \untrustedos. 
In HyperEnclave, though the enclave is still part of the application's address space, we create a separate page table for the enclave and let \visorname manage the enclave's page table and page fault without the involvement of the \untrustedos\footnote{P-Enclave can manage its own guest page table (Sec.~\ref{subsec:penclave}).},
while the page tables in the normal VM are still managed by the \untrustedos. 
However, the design faces new challenges: since the enclave can access the application's entire address space,
upon a change to the mapping of the page tables in the applications, e.g, due to page swapping, the updated mapping needs to be synchronized to the enclave's page table managed by \visorname.

To eliminate the overhead for synchronization, we pre-allocate a \textit{marshalling buffer} in the application's address space, which is shared with the enclave. The mappings of the marshalling buffer are fixed during the entire enclave life cycle by pre-populating the physical memory and pinning it in the memory. 
All data exchanged between the enclave and the application must be passed through the marshalling buffer. 
The application's memory mappings (except those for the marshalling buffer) are not needed by the enclave and are not included in the enclave's page table.
Such a design also mitigates the known enclave malware attacks~\cite{2019practical}, as the enclave cannot access the application's address space but the marshalling buffer (Sec.~\ref{sec:security} for more details).
We remind the attacker may manipulate the marshalling buffer, however it does not cause additional security issues, since the buffer is untrusted by design where the developer is responsible to ensure that the data transmitted through the buffer is authentic and protected
(same as the SGX model). 

\ignore{
when \visorname handles the page fault, it may need to evict the pages out of the enclave memory if running out of memory. To enforce the confidentiality and integrity of the paging memory, \visorname designs a paging manager that calculates the hash on the plaintext and encrypts the content and  before copy out the pages.\wenhao{we may not need to describe this}
}

When the enclave accesses a virtual address that is not committed with a physical page (e.g., due to page swapping or EDMM), a page fault is raised and the enclave traps to \visorname. \visorname picks up a free page from the enclave memory pool, inserts a new mapping to the enclave's page table, and resumes the enclave's execution. When the enclave requests changing the page permissions, the enclave issues a hypercall to \visorname to update the permissions in the enclave's page table and clear the corresponding TLB entries.\footnote{P-Enclave can change the page permissions by itself (Sec.~\ref{subsec:penclave}).}




\begin{figure}[t]
    \centering
    \includegraphics[width=0.9\linewidth]{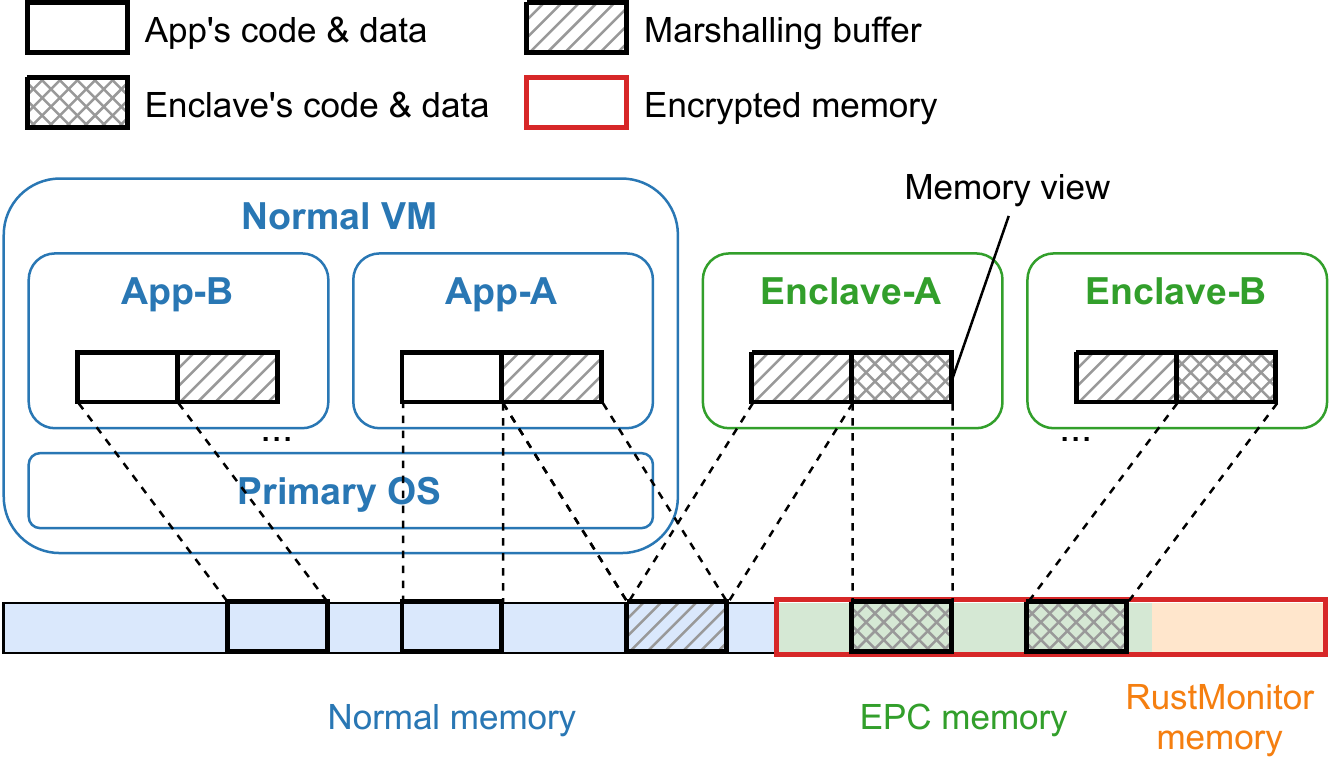}
    \caption{Memory isolation.
    }
    \label{fig:memory-view}
    \vspace*{-0.05in}
\end{figure}

\mypara{HyperEnclave memory isolation}
Figure~\ref{fig:memory-view} shows the memory mappings of the applications within the normal VM and the enclaves. The application's memory within the normal VM is managed with nested paging, while the enclave's memory could be managed through nested paging or through normal 1-level address translation, determined by the corresponding operation mode (Sec.~\ref{sec:flexibility}). As a result, HyperEnclave enforces the following security requirements.
\begin{packeditemize}
    \item \textbf{R-1:} The \untrustedos and applications are not allowed to access the physical memory belonging to \visorname and the enclaves. 
    \item \textbf{R-2:} The enclave is not allowed to access physical memory belonging to \visorname and other enclaves. It is designed to have access to only a specific memory region shared with the untrusted application for parameter passing (i.e., the marshalling buffer). 
    \item \textbf{R-3:} DMA accesses from malicious peripherals to the physical memory belonging to \visorname and the enclaves are not allowed. 
    In order to prevent such attacks, HyperEnclave restricts the physical memory used by the peripherals with the support of the Input-Output Memory Management Unit (IOMMU) in modern processors.
\end{packeditemize}

\mypara{Memory encryption}
To thwart physical memory attacks, such as cold boot and bus snooping attacks, HyperEnclave may leverage hardware memory encryption (such as AMD SME~\cite{208506} and Intel MKTME~\cite{mktme2020}) to encrypt partial physical memory at the page granularity.
If the platform does not support hardware memory encryption, HyperEnclave may consider to apply software approaches~\cite{Zhao2019SecTEEAS} to encrypt the isolated memory. This approach, however, may impose substantial overhead compared with hardware based solutions.

\subsection{Trusted Boot, Attestation and Sealing}
\label{subsec:attestation}

\begin{figure}[t]
    \centering
    \includegraphics[width=\linewidth]{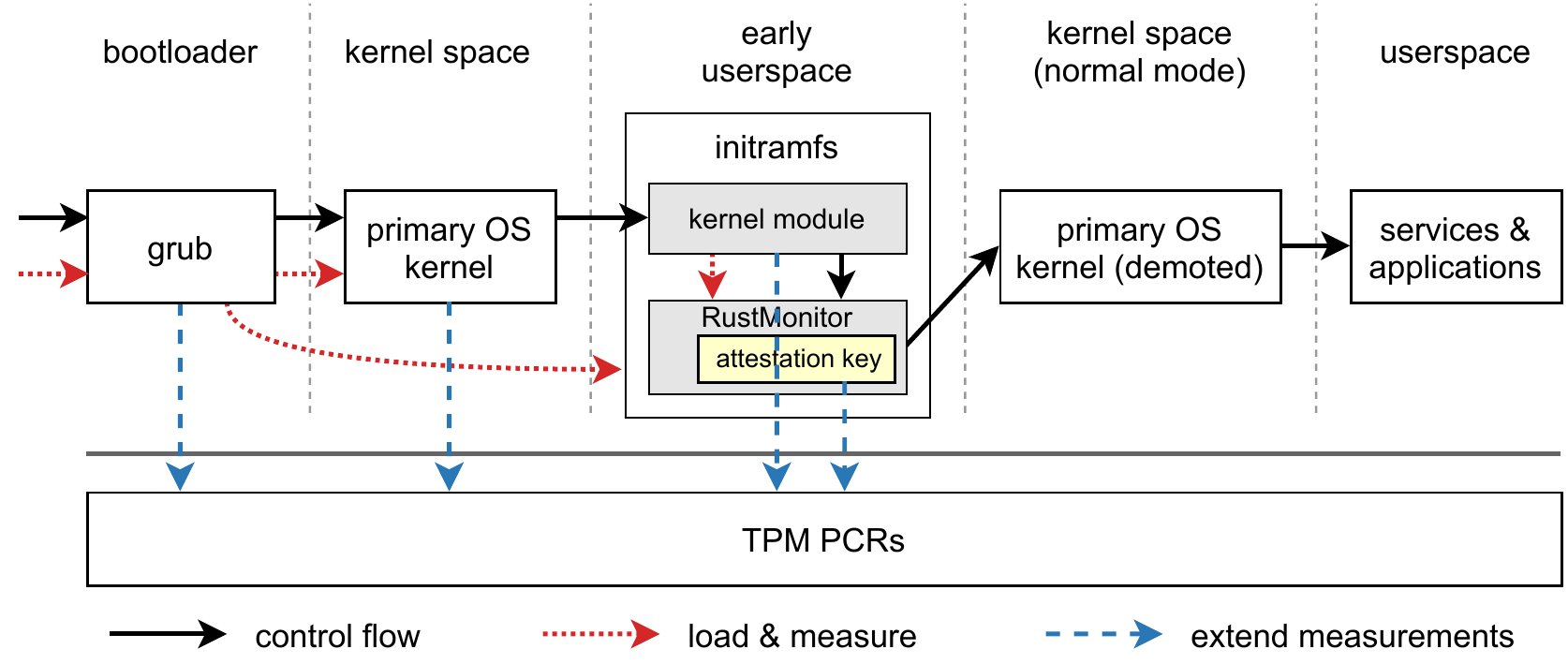}
    \caption{Measured Late Launch.}
    \label{fig:boot}
\end{figure}

\para{Measured Late Launch}
The boot process of HyperEnclave is shown in Figure~\ref{fig:boot}. On system boot, a static and immutable piece of code, known as the Core Root of Trust for Measurement (CRTM), executes first to bootstrap the process of building a measurement chain for subsequent firmware and software, including the BIOS, grub, the \untrustedos kernel, and initramfs. The measurements are stored to TPM PCRs for each boot component, so that any modification will be reflected in the attestation quote.

To reduce the attack surface from the \untrustedos, we put the \visorname image into the initramfs. 
The kernel measures the \visorname image and extends the value to TPM PCRs, then it
launches \visorname in \textit{early userspace}, i.e., before any userspace program that relies on the disk file system starts to run. 
Along with the measured boot, it ensures that the software state when \visorname is loaded is trusted.

After \visorname is loaded, the execution continues at the pre-defined entry.
\visorname sets up its own running context (such as the stack, page table, IDT, etc.) and prepares the virtual CPU (vCPU) configurations for each CPU. Then \visorname launches the normal VM and demotes the \untrustedos to the normal mode.
Returning to the kernel module, the kernel continues to boot in the normal mode and is unaware of the existence of \visorname.

HyperEnclave applies the above approach (referred to as \textit{measured late launch}) so that \visorname is loaded as a type-2 hypervisor (like KVM) while runs as a type-1 hypervisor (like Xen). 
In this way, 
\visorname does not need to trust the \untrustedos anymore after the \untrustedos is demoted to the normal mode. 


\begin{figure}[t]
    \centering
    \includegraphics[width=0.55\linewidth]{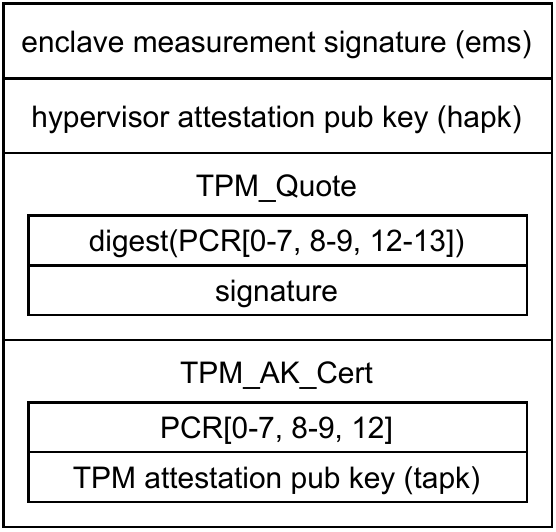}
    \caption{The HyperEnclave quote structure.}
    \label{fig:signature}
    \vspace*{-0.05in}
\end{figure}

\mypara{Remote Attestation}
With the measured late launch, all booted components are measured and extended to the TPM. After \visorname is booted, it needs to extend the trust to the enclaves. For this purpose, \visorname derives an attestation key pair which is used to sign the enclave measurement. Then \visorname extends the derived public key to the TPM PCR, and the private key never leaves \visorname which is protected by memory isolation and encryption.

During enclave creation, all pages added to the enclave (including the corresponding page content, page type, and RWX permissions) are measured by \visorname to generate the enclave measurement. The (intermediate) measurement is stored in \visorname's memory, which is invisible to the enclaves and the \untrustedos.


Similar to TPM and Intel SGX, HyperEnclave adopts a SIGn-and-MAc (SIGMA) attestation protocol for the remote attestation flow. As shown in Figure~\ref{fig:signature}, 
we denote the public key of \visorname's attestation key by the hypervisor attestation public key (\textsf{hapk}). The enclave measurement is signed using \visorname's attestation key to form the enclave measurement signature (\textsf{ems}). The TPM quote \textsf{TMP\_Quote}, which is signed using the TPM attestation key, includes the PCRs for the measurement of all booted code, and the measurement of \textsf{hapk}. Upon receiving the attestation report, the remote user can verify the report by comparing the measurement of booted code (including the CRTM, BIOS, grub, kernel, initramfs, and hypervisor) and the enclave, as well as verifying the certificate chain for generating the signature.

\ignore{
The remote attestation flow is similar to SGX, following the same SIGn-and-MAc (SIGMA) protocol. We extended the \textsf{}{sgx\_quote\_t} structure in the SDK to include the HyperEnclave quote, and the modification is transparent to the enclave code. 
}


\mypara{Secret key generation}
When \visorname is initialized for the first time, it generates a root key $K_{root}$ from the random number generator (RNG) module of the TPM. $K_{root}$ is stored outside the TPM using TPM's seal operation. During the booting process on system reset, \visorname decrypts $K_{root}$ using TPM's unseal operation, which guarantees that $K_{root}$ can only be unsealed with the exactly same TPM chip with matching PCR configurations. Furthermore, \visorname floods the PCRs with a constant before transferring control to the \untrustedos to prevent it from retrieving $K_{root}$. All other key materials, including the enclave's sealing key and report key are derived from both $K_{root}$ and the enclave's measurement.

\subsection{The Enclave SDK}
Porting existing applications to the enclaves can be cumbersome since TEEs usually expose limited hardware and software interfaces and provide additional security services (e.g., attestation and sealing). For process-based TEEs, the applications need to be partitioned into the trusted and untrusted parts, \shuang{this arguments is not true, for many process-based TEEs protect the
application as a whole. e.g.InkTag, CURE}and the interfaces need to be carefully designed to avoid various security pitfalls~\cite{checkoway2013iago,van2019tale,khandaker2020coin}. A lot of effort has been spent and many tools have been developed for Intel SGX, due to its dominant position in the market, including library OSes~\cite{tsai2017graphene,shen2020occlum}, containers~\cite{arnautov2016scone}, automatic partition and protection tools~\cite{lind2017glamdring,tsai2020civet}, WebAssembly Micro Runtime~\cite{menetrey2021twine}, and interface protection~\cite{shinde2020besfs}. Consequently, Intel SGX has supported securely running applications written in C/C++, Rust, Java, Python, etc., without expensive code refactoring.

\label{subsec:sgxprogramming}

\ignore{
\begin{table}[t]
    \caption{SGX instructions supported in HyperEnclave.}
    \label{tab:sgx-instructions}
    \footnotesize\centering
    \begin{tabular}{lll}
        \toprule
        Instruction & Leaf  & Description \\
        \midrule
        \multirow{5}{*}{\textsf{ENCLS}}
            & \textsf{ECREATE }   & Create an enclave \\
            & \textsf{EADD    }   & Add a 4K page \\
            & \textsf{EEXTEND }   & Extend enclave measurement \\
            & \textsf{EREMOVE }   & Remove a page from EPC \\
            & \textsf{EINIT   }   & Initialize an enclave \\
        \midrule
        \multirow{5}{*}{\textsf{ENCLU}}
            & \textsf{EENTER  }   & Enter an enclave \\
            & \textsf{EEXIT   }   & Exit an enclave \\
            & \textsf{ERESUME }   & Re-enter an enclave \\
            & \textsf{EGETKEY }   & Create a cryptographic key \\
            & \textsf{EREPORT }   & Create a cryptographic report \\
        \bottomrule
    \end{tabular}
\end{table}
}

We provide the enclave SDK with APIs compatible with the official Intel SGX SDK to ease the development of applications on HyperEnclave.
The enclave SDK is retrofitting the official SGX SDK. By replacing the SGX user leaf functions (e.g., \textsf{EENTER}, \textsf{EEXIT}, and \textsf{ERESUME}) with hypercalls,
SGX programs can run on HyperEnclave with little or no source code changes. Once the enclave executes these user leaf functions, it traps to \visorname and \visorname emulates the functionalities of the corresponding SGX instructions. 

The enclave is compiled as a trusted library of the application, while the application itself runs in the \untrustedos. The enclave life cycle is managed through the emulation of a set of privileged SGX instructions (i.e., \textsf{ECREATE}, \textsf{EADD}, \textsf{EINIT}, etc.). To this end, the kernel module running in the \untrustedos provides similar functionalities by invoking \visorname through hypercalls, and exposes the functionalities to the applications by the \texttt{ioctl()} interfaces. By emulating the privileged SGX instructions, \visorname is responsible for the management of the enclave's life cycle (Sec.~\ref{sec:flexibility}).

To be compatible with the official Intel SGX SDK, most data structures involved in HyperEnclave (such as the SIGSTRUCT structure, the SECS page, and the TCS page) are similar to that of SGX. 
With the HyperEnclave design, it is straightforward to support dynamic enclave management in an enclave, since the enclave memory and page fault are all managed by \visorname. Multi-threading within the enclave is supported by associating one TCS page for each enclave thread within the enclave. Exception handling within the enclave is supported by setting more than 1 SSA page for each TCS. The details are omitted due to space constraints and we refer the readers to the SGX manual~\cite{IntelDevelopmentManual} for more details.

\section{Flexible Enclave Operation Mode}
\label{sec:flexibility}

\begin{figure}[t]
    \centering
    \includegraphics[width=0.95\linewidth]{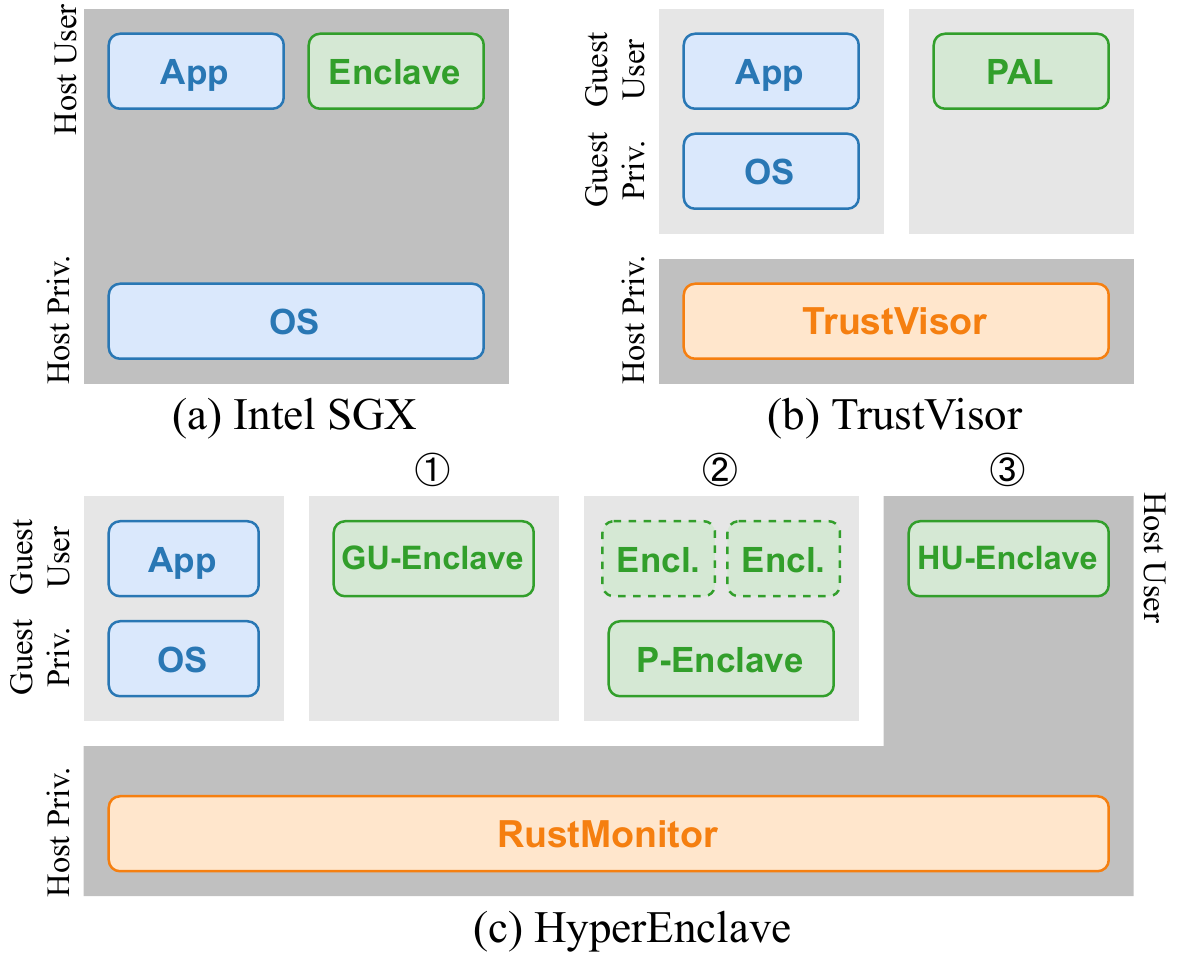}
    \caption{Comparison of the enclave operation modes  supported by process-based TEEs. (a) Intel SGX runs enclaves in the host user mode (or guest user mode in the virtualization environment). (b) TrustVisor runs the protected code (Pieces of Application Logic, PALs) in the guest user mode. (c) HyperEnclave supports 3 coexisting enclave operation modes: \circled{1} GU-Enclaves running in guest user mode; \circled{2} P-Enclaves running in guest privileged mode
    and optional guest user mode; \circled{3} HU-Enclaves running in host user mode.}
    \label{fig:enclave-modes}
\end{figure}

\begin{figure}[t]
    \centering
    \begin{subfigure}[b]{0.49\linewidth}
        \centering
        \includegraphics[width=0.9\linewidth]{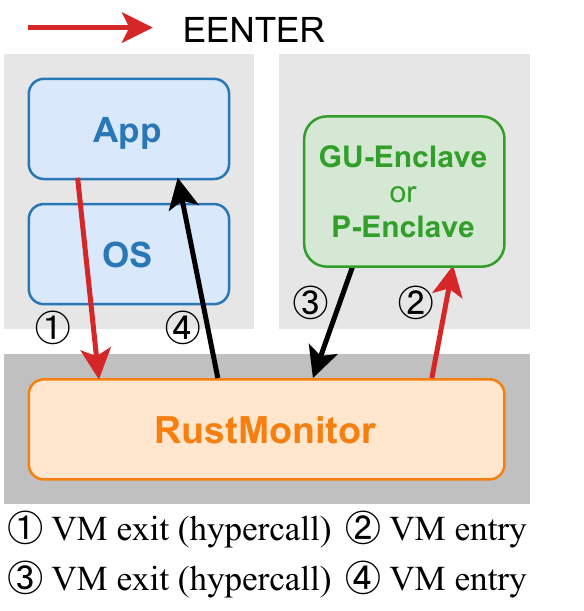}
        \caption{GU-Enclave and P-Enclave}
        \label{fig:world-switch-1}
    \end{subfigure}
    \hfill
    \begin{subfigure}[b]{0.49\linewidth}
        \centering
        \includegraphics[width=0.9\linewidth]{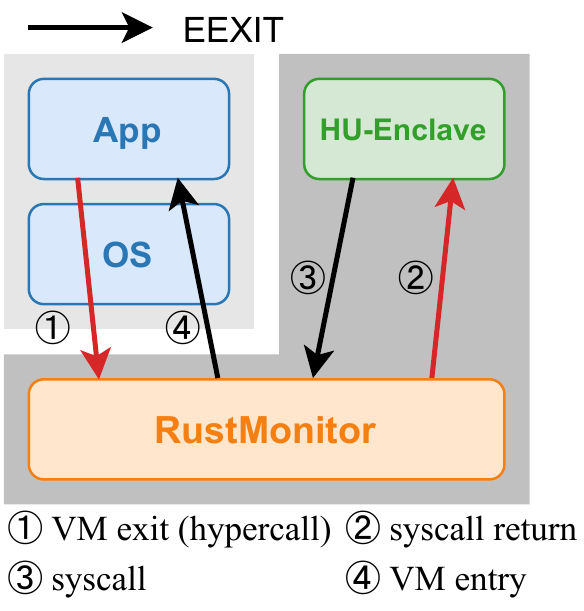}
        \caption{HU-Enclave}
        \label{fig:world-switch-2}
    \end{subfigure}
    \caption{World switches for the supported enclave operation modes.
    (a) Using hypercalls to enter and exit GU-Enclave and P-Enclave. (b) Using syscalls and syscall returns to enter and exit HU-Enclave.}
    \label{fig:world-switch}
\end{figure}

A wide range of existing applications can be offloaded to the TEEs, such as computing-intensive tasks (machine learning~\cite{ohrimenko2016oblivious}), input and output (IO)-intensive tasks (such as the Apache and Nginx web server~\cite{arnautov2016scone}), memory-intensive tasks (Redis and Memcached~\cite{arnautov2016scone}), and tasks which favor in-enclave exception handling and privilege separation~\cite{dune}.
Most TEEs support running the enclaves only in fixed mode, Intel SGX (also TrustVisor~\cite{McCune2010TrustVisorET} and Secage~\cite{liu2015thwarting}) enclaves in particular, as part of the application address space, run in user mode. As a result, the user mode enclave is not allowed to access the privileged resources (such as the IDT and page tables) and process the privileged events (interrupt and exceptions). It must switch to the untrusted code to gain access to privileged resources and handle the events. The I/O-intensive and memory-intensive tasks essentially involve the frequent world switches which are expensive and introduce non-negligible performance losses, even though both software and hardware optimizations have been proposed trying to reduce the context switch latencies~\cite{orenbach2017eleos,tian2018switchless,weisse2017regaining}.
In this section, we introduce the three enclave operation modes supported by HyperEnclave, as shown in Figure~\ref{fig:enclave-modes}. The world switches in different enclave operation modes are shown in Figure~\ref{fig:world-switch}. 

\subsection{Guest User Enclaves}
\label{subsec:guestue}
Guest user enclave (GU-Enclave) is the basic enclave operation mode which is typically running computing-intensive tasks. The enclave runs in the guest user mode (i.e., guest ring-3 of the VMX non-root operation mode).

During the enclave creation, \visorname prepares a vCPU structure which contains a guest page table (GPT) and a nested page table (NPT) for GU-Enclave. On entry and exit between the normal VM and the enclave VM, \visorname
switches the vCPU states (e.g. the instruction pointer, thread pointer, NPT, and GPT) accordingly. 


To handle the interrupts and exceptions during the enclave running,
\visorname configures the vCPU to trap all interrupts and exceptions to the \host. \visorname then saves the enclave's context,
forwards the interrupt or exception to the normal VM. After the \untrustedos completes handling the interrupt or exception, the application invokes the \textsf{ERESUME} hypercall, which traps to
\visorname to restore the enclave's context 
and resume the execution of the enclave.

\subsection{Host User Enclaves}

Host user enclave (HU-Enclave) is running in host user mode. It delivers the optimal world switch efficiency by substituting the mode switch (hypercalls: $\sim 880$ CPU cycles on our platform) with the ring switch (syscalls: $\sim 120$ CPU cycles on our platform) (Figure~\ref{fig:world-switch}). It further eliminates the extra virtualization overhead (e.g. vCPU context switching and two-dimensional page walking) in GU-Enclave. HU-Enclave may benefit the I/O-intensive workload according to our evaluation in Sec~\ref{sec:eval}. By comparison, running enclaves in the guest user mode provides more defensive depth. \wenhao{Specifically a malicious application that exploits a bug in the GU-Enclaves only gains access to a restricted user-space process; in contrast, a process escaping a HU-Enclave might gain root access.}



When loading the HU-Enclave, \visorname prepares a process context, e.g. creates a level-1 page table. 
On enclave entry, \visorname updates the CPU state and invokes the system call return instruction (i.e., \texttt{SYSRET} on x86 platforms) to enter the HU-Enclave. Correspondingly, on enclave exit, HU-enclave invokes the system call instruction (i.e., \texttt{SYSCALL} on x86 platforms) and traps into \visorname. 
The \textsf{ENCLU} leaf instructions 
(e.g., \textsf{EGETKEY}, \textsf{EREPORT}) are emulated as a system call.
Interrupts and exceptions within the HU-Enclaves also trap into the \visorname. The procedures are similar to those for the GU-Enclaves described in Sec.~\ref{subsec:guestue}.



\subsection{Privileged Enclaves}
\label{subsec:penclave}

Inspired by the VM-based TEEs, such as AMD SEV~\cite{kaplan2016amd}, HyperEnclave supports privilege enclaves (P-Enclaves) which run in guest privileged mode. 
P-Enclave is permitted to access the GDT, IDT, and level-1 page table which benefits a wide variety of applications, as demonstrated by Dune~\cite{dune}. One such example is the garbage collector, an essential feature for Java applications (existing works port the JVM to enclaves~\cite{Jiang2020UranusSE,Tsai2020CivetAE}). The garbage collector frequently changes page permissions to trigger page faults in order to track the page status. 
For user mode enclaves (e.g., GU-Enclaves and HU-Enclaves), it has to involve the \untrustedos to update the page table and handle the page fault which suffers huge performance loss due to world switches. 
P-Enclaves eliminate the world switch by supporting in-enclave exception handling and level-1 page table management.
More specifically, P-Enclaves configures its own exception handler to handle certain exceptions (such as page fault). \visorname passes through the white-list exceptions to the P-Enclave and forwards others to the \untrustedos. 
Furthermore, P-Enclaves can also support page-table-based in-enclave isolation schemes, e.g., sandboxing untrusted third-party libraries.




With the ability to receive interrupts within the enclaves, P-Enclaves may also detect abnormal interrupt events by counting the frequency, before requesting \visorname to route them to the \untrustedos. As such, existing interrupt-based side channel attacks~\cite{brasser2017software,moghimi2017cachezoom,hahnel2017high,van2018nemesis,huo2020bluethunder,moghimi2020copycat} could be detected and mitigated. We leave further exploration in this direction to future work due to space constraints.

\section{Implementations}
We report our implementation of HyperEnclave on an AMD platform that supports hardware virtualization technology and memory encryption. In the current implementation, \visorname consists of about 7,500 lines of code written mostly in Rust, and the kernel module for the \untrustedos has about 3,500 lines of C code. Also, we made about 2,000 lines of code changes to the official Intel SGX SDK (version 2.13).



\subsection{\visorname}
\visorname runs at the highest privilege level and enforces the isolation for the enclaves.
To reduce the risks caused by memory corruption or concurrency bugs, we implemented \visorname mostly in Rust, a memory-safe language, with only a few lines of assembly code used for context switches. Compared with existing hypervisors such as KVM~\cite{kivity2007kvm} and Xen~\cite{chisnall2008definitive}, \visorname is much smaller and thus easier to be formally verified. We are working on the formal verification of \visorname and plan to release the result as a separate report.

When the platform is booted, we configure the
kernel command line parameters in the grub to reserve regions of physical memory, which are exclusively used by \visorname and the enclaves.
\visorname manages the reserved physical memory by maintaining a list of free pages. When an enclave page is needed, e.g., when adding an enclave page during enclave creation, a free page is retrieved from the pool; when the enclave page is freed, the page is attached to the list again. Moreover, \visorname also manages the enclave's page tables and processes the page fault.
\subsection{The Kernel Module}
\label{subsec:kernelmodule}


The kernel module is loaded by the \untrustedos during the booting process. Then it loads, measures, and launches \visorname, with the measurement extended to the TPM PCR as part of the TPM quote.
When the kernel module is loaded, a device file is created and mounted at \path{/dev/hyper_enclave}. The application can open it and issue the \texttt{ioctl()} to invoke the emulated privileged operations.

\subsection{The Enclave SDK}
\label{subsec:enclavesdk}

{HyperEnclave retrofits the official SGX SDK as follows.}


\mypara{Supporting the SGX SDK APIs}
We replace the SGX user leaf functions (e.g. \textsf{EENTER}, \textsf{EEXIT}, \textsf{ERESUME}, etc.) in the SGX SDK with hypercalls or system calls. Our implementation retains the same parameter semantics and orders as SGX for compatibility purposes.


\mypara{Parameters passing with the marshalling buffer}
In HyperEnclave, the enclave can only access its own address space and the marshalling buffer shared with the application. The size of the marshalling buffer can be configured in the enclave's configuration file, with a default size. The data needs to be transmitted to the marshalling buffer before invoking edge calls.  We modified SGX SDK to handle the transitions, which are thus transparent to the developer.

\ignore{To prevent the mapping attack from the untrusted guest OS, the enclave's gPT is managed by the hypervisor (details in Sec.~\ref{sec:defense-against-mapping-attack}). In our design, the enclave can only access its own EPC memory and the marshalling buffer. The size of the marshalling buffer can be configured in the enclave's configuration file, with a default size of 8 KB. The data needs to be transmitted to the marshalling buffer before the parameter passing of edge calls (i.e., \textsf{ECALLs} and \textsf{OCALLs}).  We modified the SGX SDK to handle the transitions, which are thus transparent to the developer.
}

We modified the untrusted runtime library in the SDK (i.e., \path{libsgx_urts.so}), such that during enclave initialization a marshalling buffer is allocated using \path{mmap()} with \path{MAP_POPULATE} flags set. As a result, the GPAs for the marshalling buffers are pre-populated. Then an \path{ioctl()} is issued to request the \untrustedos not to compact or swap out the physical pages of the marshalling buffers during the enclave's lifetime.
When the application invokes the emulated \textsf{EINIT} instruction to mark the initialization of the enclave, the base address and the size of the marshalling buffer are passed to \visorname, who will add the mapping of the marshalling buffer in the enclave's page table. In this way, the marshalling buffer is now shared between the enclave and the untrusted application. The base address and the size of the marshalling buffer are also passed to the trusted runtime library to transmit data from the marshalling buffer to the enclave.





The current OCALL's implementation in the SGX SDK invokes the \path{sgx_ocalloc()} within the enclave to allocate a buffer on the stack area of the untrusted application, which is then used for cross-enclave data transmission. As such, we only need to modify the \path{sgx_ocalloc()} function to allocate a memory area in the marshalling buffer. To support parameter passing through the marshalling buffer for ECALLs, we modified SGX's \textsf{Edger8r} tool to automatically generate code that copies the transmitted data into the marshalling buffer.



The SGX programming model supports passing parameters with the \path{user_check} attribute. For such parameters, the SDK tool will not generate code to check the address range or perform data movement. Since the enclave code could access the entire process's address space, some enclave programs may use a pointer with the \path{user_check} attribute to manipulate the data buffer outside the enclave directly, without accounting for the overhead for copying the data across the enclave boundary. To deal with it, we added an interface for the developer to allocate the buffer within the marshalling buffer, in the cases when the developer may use parameters with the \path{user_check} attribute.

The remote attestation flow is similar to SGX, following the same SIGn-and-MAc (SIGMA) protocol. We extended the \texttt{sgx\_quote\_t} structure in the SDK to include the HyperEnclave quote, and the modification is transparent to the enclave code.


With the above design, most SGX programs could run on HyperEnclave without source code changes. Furthermore, {{to ease the development of HyperEnclave applications, we have also ported the Rust SGX SDK~\cite{Wang2019TowardsMS} and the Occlum library OS~\cite{shen2020occlum} to HyperEnclave.}}






\section{Security Analysis}
\label{sec:security}

\para{Trust Establishment}
HyperEnclave relies on measured boot to bootstrap the trust of \visorname, a common approach for the design of TEEs (e.g., TrustZone~\cite{2004Trustzone} and Keystone~\cite{lee2020keystone}). All the components during booting (including the CRTM, BIOS, grub, kernel, and initramfs) are measured and extended to the TPM PCRs. As a result, any tampering of the booted code will be reflected and audited through remote attestation.

We consider HyperEnclave is deployed in a controlled environment (i.e., the data center in the cloud computing scenario) such that the attacker has limited physical access to the platforms. To reduce the attack surface and minimize the TCB, we put the \visorname image into the initramfs, and \visorname is loaded and measured in early userspace.
In this phase, the \untrustedos kernel does not accept external inputs from the user, and the peripherals such as network connection are disabled. With the measured late launch approach, \visorname does not need to trust the \untrustedos anymore after the OS is demoted to the \normal.




\mypara{Enclave memory isolation}
As presented in Sec.~\ref{sec:enclave-memory-protection}, the enclave's memory and page tables are maintained by \visorname, which are inaccessible to the \untrustedos. The TLBs are cleared upon world switches to prevent illegal memory accesses using stale TLB entries.
\visorname prevents the \untrustedos to access the reserved physical memory by removing the corresponding mappings from its NPT.
\visorname also configures the IOMMU to prevent unauthorized device accesses to the reserved physical memory.


Our design prevents the memory mapping attacks since the \untrustedos cannot interfere with the enclave's address mappings. We introduce the marshalling buffer to support memory sharing between the enclave and the application. The application pre-allocates a marshalling buffer in normal memory and passes the base address and size of the buffer to \visorname during enclave initialization.
In case the application may pass crafted addresses (e.g., to overwrite the enclave memory), before adding the mapping of the marshalling buffer to the enclave's page table, \visorname ensures 
the address range of the marshalling buffer is outside the enclave address range.

\mypara{Defense Against Compromised Enclaves}
Previous work demonstrates that enclave malware may steal secret data or hijack the control-flow of the application outside the enclave~\cite{2019practical}. HyperEnclave is designed to confine potentially malicious enclaves as follows.
\begin{packeditemize}
    \item \textit{Preventing arbitrary memory accesses to the application.} The SGX enclave could access the entire address space of the application, and it is possible for attacks such as leaking the secret keys or stack canaries, or tampering with the code pointers for control flow attacks. In HyperEnclave, the enclave can only access its own memory and the marshalling buffer, which is only used for parameter passing.
    \item \textit{Preventing arbitrary control flows after EEXIT.}
    \atc{The SGX design allows the enclave to jump to arbitrary addresses by setting \textsf{rbx} before executing the EEXIT instruction (i.e., exiting the enclave), opening door to enclave malware attacks~\cite{2019practical}.}
    In our design, since the \textsf{EEXIT} instruction is emulated by \visorname, it is easy to prevent such attacks by adding the validity check when \textsf{EEXIT} is invoked.
\end{packeditemize}

\noindent\textbf{Physical attacks.}
Hardware memory encryption techniques such as AMD SME could be used to protect the enclave from physical attacks such as cold boot attacks or bus snooping attacks. With memory encryption, data are always encrypted in the memory or on the memory bus, and are only decrypted within the CPU. The memory encryption key is generated randomly on system boot and stored in the CPU, which cannot be accessed explicitly by software.

\mypara{Side channel attacks}
Compared with SGX, HyperEnclave can mitigate certain types of side channel attacks. Since the enclave's guest page tables and page fault events are processed by \visorname without \untrustedos involvement, the latter cannot mount page-table-based attacks~\cite{xu2015controlled,wang2017leaky,van2017telling}. We leave the protection against micro-architectural attacks such as speculative execution attacks as future work.



\section{Evaluation}
\label{sec:eval}

We deployed HyperEnclave on a server with two 
{AMD EPYC 7601} CPUs (2 threads per core, total of 128 logical cores)
with
512 GB DDR4 RAM. We configure 2 GB reserved memory for \visorname, and 24 GB for EPC memory. The \untrustedos is Ubuntu 18.04 LTS with Linux kernel 4.19.91.
For comparison, we run the same experiments on an Intel Xeon E3-1270 v6 CPU with SGX enabled, with 64 GB DDR4 RAM, running the same OS.
The SGX SDK version for both HyperEnclave and the native SGX hardware are 2.13. All programs are compiled with GCC 7.5.0 and the same optimization level. 

\atc{We tried to rule out differences between hardware. Except where explicitly stated, the evaluations didn't exceed the EPC size, so as not to trigger excessive page swapping. All evaluations were performed in single-threaded mode.}
\atc{For micro-benchmarks (Table~\ref{tab:primitive} and Table~\ref{tab:exception}), we compare HyperEnclave on AMD hardware with SGX on Intel hardware using the same SGX SDK. We measured the core cycles to avoid the influence of CPU frequencies. For real-world workloads, we set the baselines as the counterparts with no security protections on Intel and AMD platforms respectively, and compare the \textit{relative slowdowns} introduced by SGX and HyperEnclave. Since we only compare the relative slowdowns, we stress that the \textit{absolute} performance results are on dissimilar platforms and are not directly comparable. All HyperEnclave evaluations were measured with memory encryption enabled.}

\atc{We've been careful in ensuring HyperEnclave implements the TEE functionality correctly. Still, memory encryption between SGX1 and HyperEnclave is different, i.e., Merkel tree and AES-CTR versus AES-XTS (see Figure~\ref{fig:memory-overhead} in Sec.~\ref{subsec:memory-overhead} for the evaluation of memory encryption overhead), which may explain the improvement for memory-intensive workloads. Besides, world switches for HyperEnclave (especially, HU-enclave) are faster, which explains the improvement for I/O-intensive workloads.}


\ignore{
我们的实验可以分为 micro-benchmarks (i.e. MEE overhead, SGX primitives performance, marshalling buffer overhead) 和 macro-benchmarks (virtualization overhead, real-world workloads overhead) 两类。对于 micro-benchmarks, 我们使用 \textsf{RDTSCP} (Read Time-Stamp Counter and Processor ID) 指令，测量一部分代码片段运行的时钟周期数。但是对于 Intel SGX 硬件，不支持在 enclave 中使用该指令，我们就通过一个专门的 OCALL 切换到非 enclave 模式去执行该指令，最后把结果减去一次空的 OCALL 的执行时间。不过对于 HyperEnclave，我们通过在 hypervisor 中配置 VMCB (or VMCS for Intel processors)，使得可以 enclave 中使用 \textsf{RDTSCP}。我们对每个 micro-benchmarks 都运行足够多的次数，并取中位数作为最终结果。
}

\ignore{
For micro-benchmarks,
we use the \textsf{RDTSCP} instruction to measure the latency on both platforms. To measure the latency for individual emulated instructions, we configure \visorname to enable the VMs to use the \textsf{RDTSCP} instruction.
}

\subsection{World Switches Performance}
\label{subsec:primitive-overhead}

\begin{table}[t]
    \small\centering
    \begin{tabular}{r|rr|rr}
        \toprule    & \textsf{EENTER}   & \textsf{EEXIT}    & ECALL     & OCALL     \\
        \midrule
        Intel SGX  & -                 & -                 & 14,432    & 12,432    \\
        \midrule
        HU-Enclave  & 1,163             & 1,144             & 8,440     & 4,120     \\
        GU-Enclave  & 1,704             & 1,319             & 9,480     & 4,920     \\
        P-Enclave   & 1,649             & 1,401             & 9,700     & 5,260     \\
        \bottomrule
    \end{tabular}
     \vspace{-0.04in}
    \caption{Latency of SGX primitives on HyperEnclave and Intel (in CPU cycles). }
    \label{tab:primitive}
    \vspace{-0.04in}
\end{table}

\ignore{
由于 SGX 不支持在 enclave 里执行 \textsf{RDTSCP} 指令，难以测量指令级别的 SGX 原语延迟，而是只能得到函数级别的 round-trip latency (i.e, ECALLs and OCALLs)。But thanks to our virtualization based design, we can easily measure more fine-grained primitives latency (i.e, \textsf{EENTER} and \textsf{EXIT}).
}

We measured the latency of edge calls (i.e., ECALLs and OCALLs) on both HyperEnclave (under different enclave operation modes) and Intel SGX. The test code runs empty edge calls with no explicit parameters 1,000,000 times and takes the median value. We also measured the instruction-level latency for the emulated \textsf{EENTER} and \textsf{EEXIT} instructions on HyperEnclave. We were not able to measure the instruction-level latency on SGX since the \texttt{RDTSCP} instruction is not supported within the enclaves on our SGX platform.



The results are shown in Table~\ref{tab:primitive}. It shows that HU-Enclave has the optimal edge calls performance as it reduces a mode switch ($\sim 880$ cycles) to a ring switch ($\sim 120$ cycles) while P-Enclave is slower than GU-Enclave for it needs to switch more privileged states during the world switches. All of the results are comparable with Intel SGX.

\begin{table}[t]
    \small\centering
    \begin{tabular}{l|rrr}
        \toprule
                        & Intel SGX     & GU-Enclave    & P-Enclave \\
        \midrule
        \texttt{\#UD}   & 28,561        & 17,490        & 258        \\
        \texttt{\#PF}   & --            & 2,660         & 1,132      \\
        \bottomrule
    \end{tabular}
    \vspace*{-0.05in}
    \caption{Average CPU cycles of handling an \texttt{\#UD} and \texttt{\#PF} exception inside the enclaves.}
    \label{tab:exception}
     \vspace*{-0.13in}
\end{table}

\subsection{Enclave Exception Handling}
\label{subsec:exception-overhead}


We used the undefined instruction exception (\texttt{\#UD}) and the page fault exception (\texttt{\#PF}) to evaluate the enclave exception handling performance.
In the \texttt{\#UD} benchmark, the test code executes an undefined instruction in the enclave to trigger the exception 1,000,000 times. The exception handler advances the instruction pointer and returns. 
For P-Enclave, the exceptions are captured and handled entirely within the enclaves, without enclave mode switches. For GU-Enclave and SGX, an exception causes an asynchronous enclave exit (AEX) and switches the CPU to the untrusted OS, then executes a two-phase exception handling~\cite{sgxtwophase}.
The result shows that the exception handling within P-Enclaves is about 68$\times$ and 110$\times$ faster than GU-Enclave and Intel SGX respectively (Table~\ref{tab:exception}).

We further simulated a typical garbage collector (GC) scenario that the test code first allocated a large memory buffer, then the write permissions to the buffer were revoked by changing the enclave's page table. After that, the enclave accessed the buffer to trigger the page faults. In the exception handler, the write permission is restored.
The result (Table~\ref{tab:exception}) shows that P-Enclave is about 2.3$\times$ faster than GU-Enclave, for P-Enclave updates the page table and handles the page faults by itself, while GU-Enclave needs to trap into \visorname to update the page tables. 
Note that we did not evaluate GC on Intel SGX, since our SGX1 platform does not support page permission modifications after the enclave initialization.

\begin{figure}[t]
    \centering
    \includegraphics[width=0.95\linewidth]{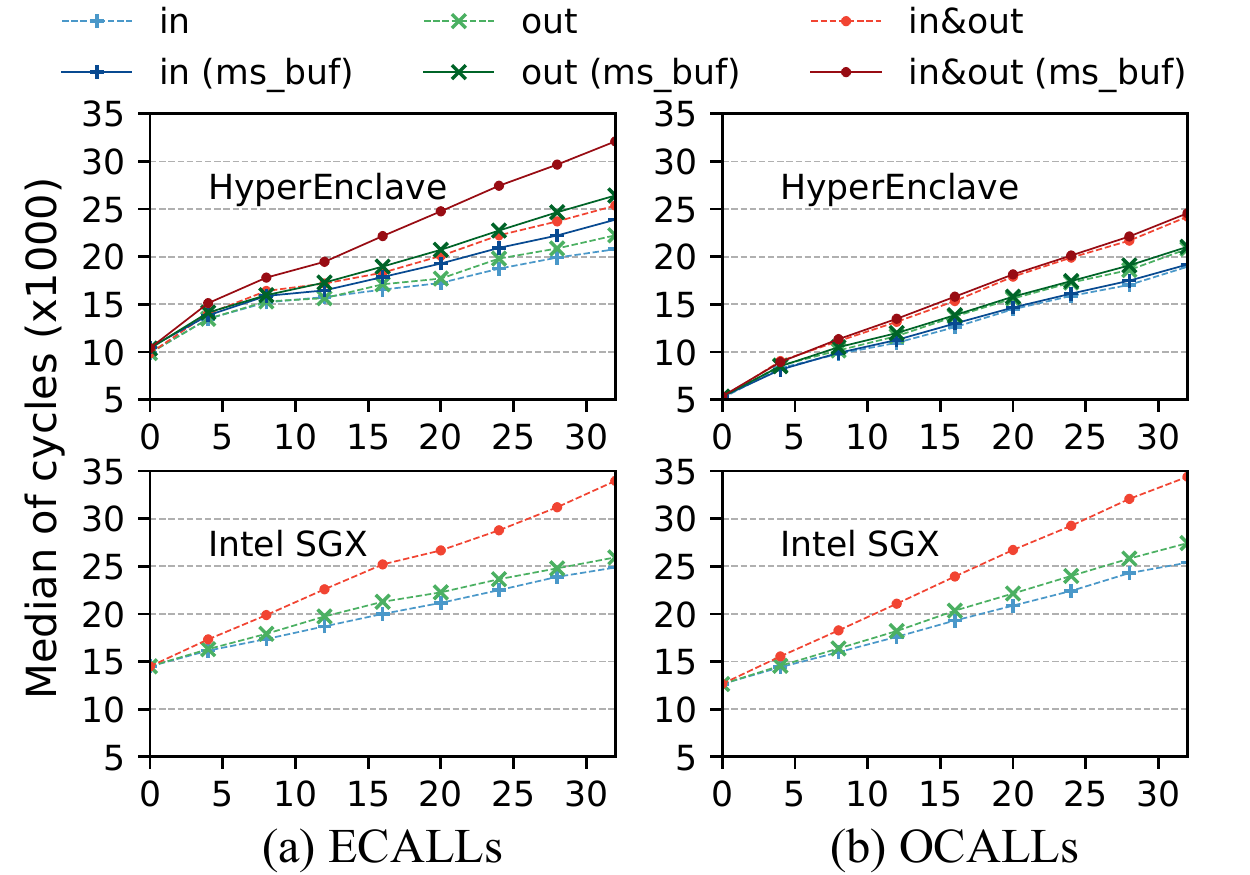}
    \caption{Marshalling buffer overhead for ECALLs and OCALLs with various data size, and with the marshalling buffer (marked as \textsf{ms\_buf}) enabled and disabled.
    }
    \label{fig:edge-calls}
     \vspace*{-0.12in}
\end{figure}

\begin{figure*}
    \centering
    \begin{subfigure}[b]{0.24\textwidth}
        \centering
        \includegraphics[width=\linewidth]{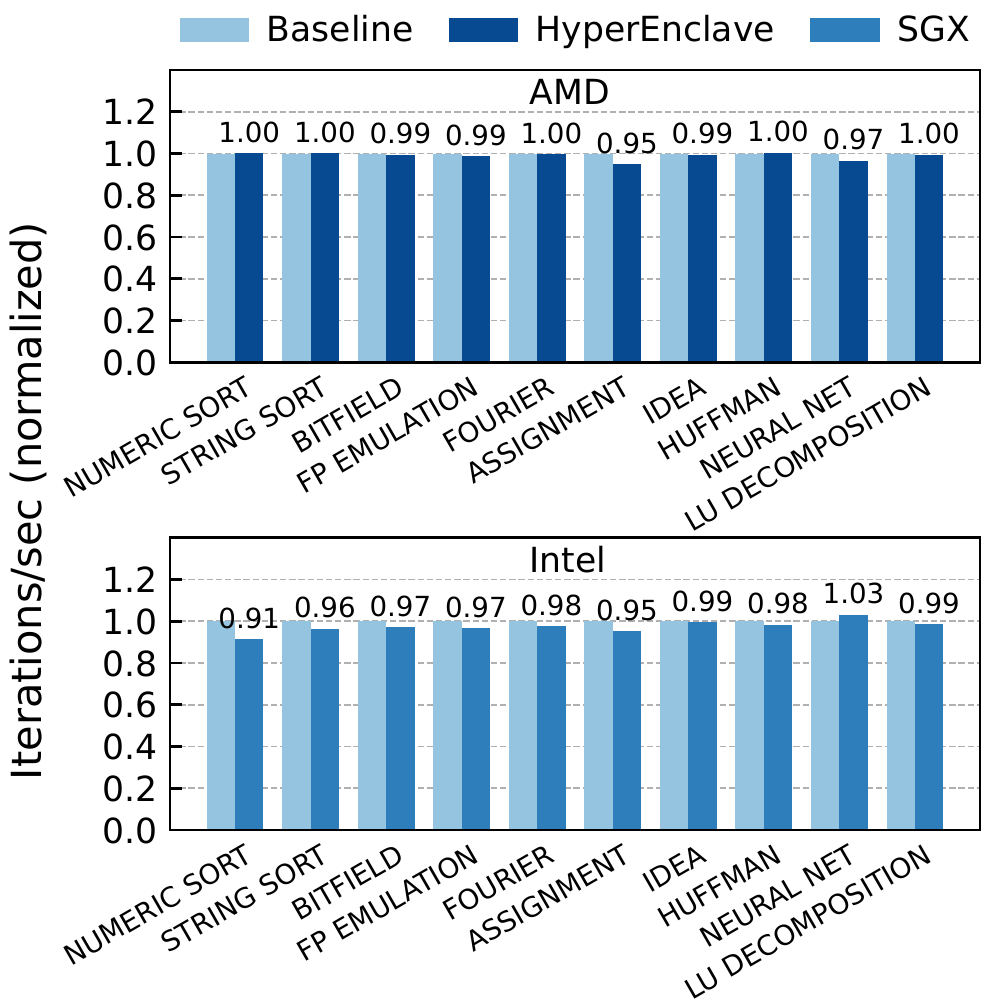}
        \caption{NBench}
        \label{fig:nbench}
    \end{subfigure}
    \hfill
    \begin{subfigure}[b]{0.24\textwidth}
        \centering
        \includegraphics[width=\linewidth]{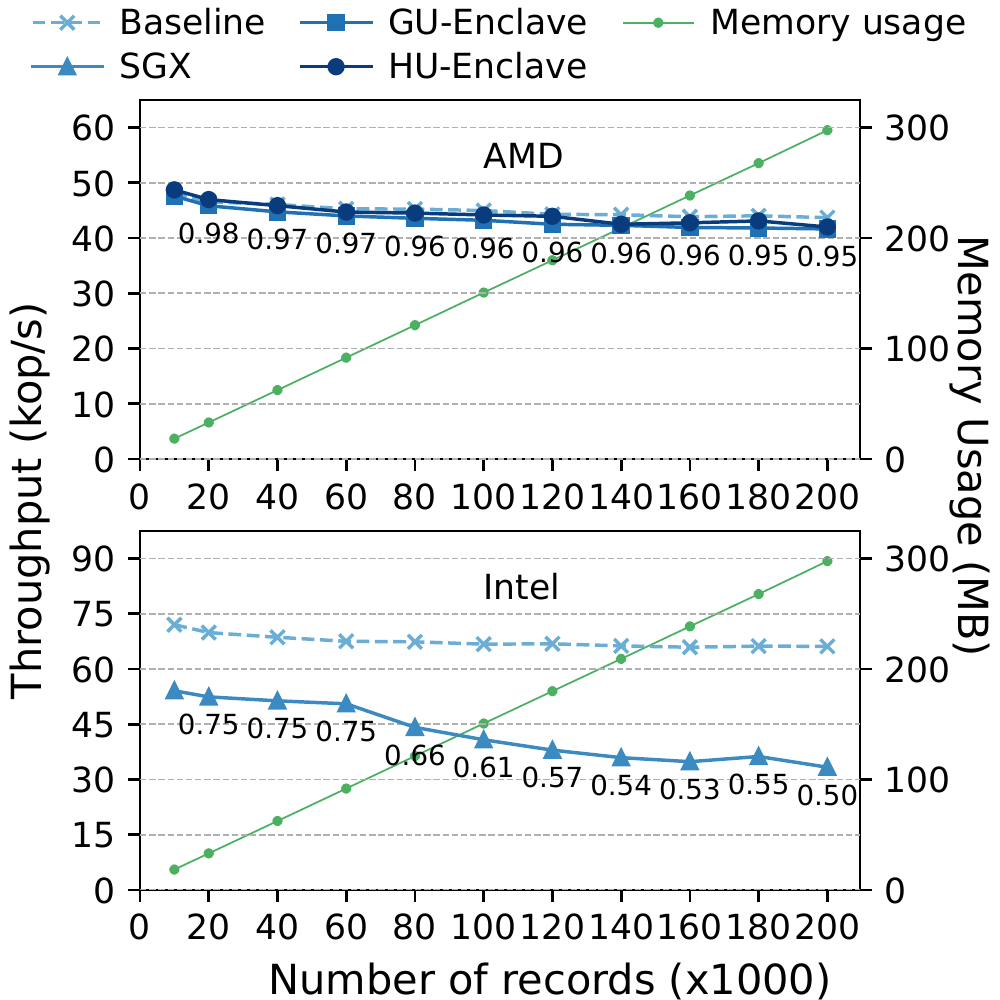}
        \caption{SQLite}
        \label{fig:sqlite}
    \end{subfigure}
    \hfill
    \begin{subfigure}[b]{0.24\textwidth}
        \centering
        \includegraphics[width=\linewidth]{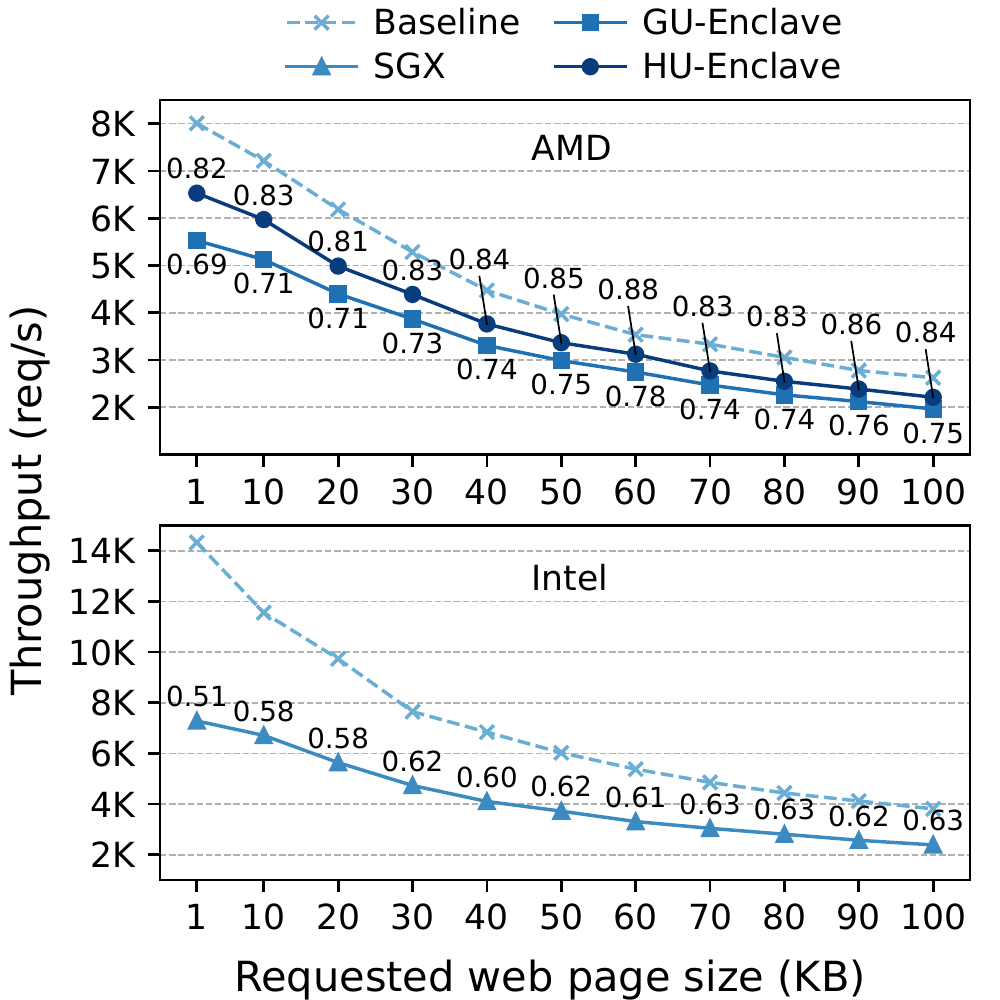}
        \caption{Lighttpd on Occlum}
        \label{fig:lighttpd}
    \end{subfigure}
    \hfill
    \begin{subfigure}[b]{0.24\textwidth}
        \centering
        \includegraphics[width=\linewidth]{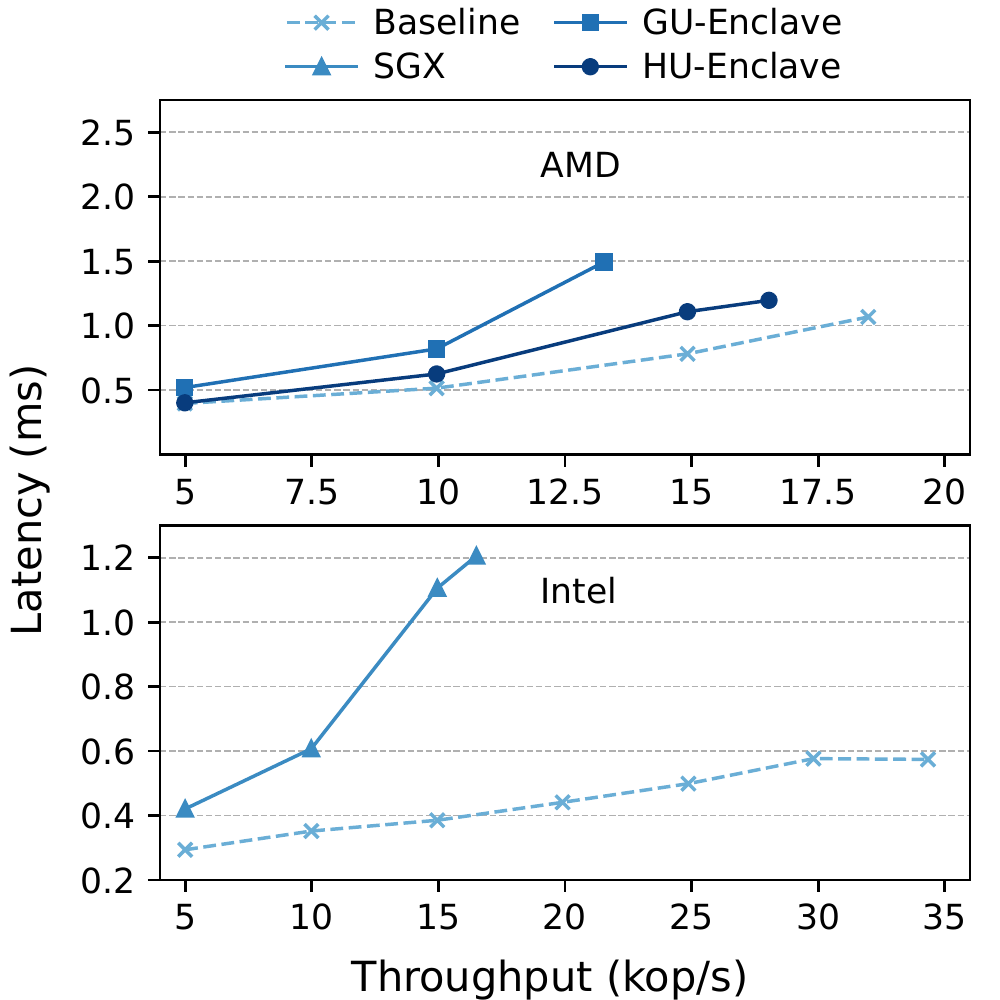}
        \caption{Redis on Occlum}
        \label{fig:redis}
    \end{subfigure}
    \vspace*{-0.05in}
    \caption{Performance of NBench, SQLite, Lighttpd, and Redis on AMD (with HyperEnclave) and Intel (with SGX).}
    \vspace*{-0.14in}
\end{figure*}

\subsection{Marshalling Buffer Overhead}
\label{subsec:msbuf-overhead}

\ignore{
We measure the overhead of both ECALLs and OCALLs with various sizes of the marshalling buffer. We also evaluated situations while varying the directions for data movement (i.e., in, out and in\&out), and compare the result with Intel SGX in Figure~\ref{fig:ecalls} and Figure~\ref{fig:ocalls}. It shows that the overhead is almost linear with the buffer size for data movement, about xx cycles and xx cycles every 16 KB data for HyperEnclave and SGX. HyperEnclave is xx\% slower due to the additional memory copy involved. We stress that the overhead is acceptable in real world use-cases (see below), and the design brings additional protection against malicious enclave programs (Sec.~\ref{sec:defense-against-mapping-attack}).

为了测量 marshalling buffer 的开销，我们专门构建了一个 HyperEnclave 的变体系统，能在 enclave VM 中访问不可信应用的数据，从而在边缘调用传递参数时，使用原始 SGX SDK 的方法而不是将数据传输到 marshalling buffer。然后，我们在边缘调用(ECALLs \& OCALLs)时传入不同大小的 buffer 作为参数，分别测量在这两个系统上进行一次完整调用的运行时间。We also evaluated situations while varying the directions for data movement (i.e., in, out and in\&out)。在每次运行之前，我们对相关地址使用 \textsf{CLFLUSH} 指令，将整个 buffer 从 cache 中清除。此外，我们在 Intel SGX hardware 上也进行了该实验，并将结果作为参考。
}

To measure the overhead of introducing the marshalling buffer, we constructed a GU-Enclave variant
that does not use the marshalling buffer as the baseline. We measured the overhead for both ECALLs and OCALLs with various sizes of the transferred data while varying the directions for data movement (i.e., ``in'', ``out'' and ``in\&out''). We ensure the data to be transferred is not cached using the \texttt{CLFLUSH} instruction. We evaluated the performance for transferring the same data on SGX for comparison.

Figure~\ref{fig:edge-calls} provides the results for ECALLs and OCALLs respectively. It shows that the overhead increases almost linearly with the data size. For ECALLs, the overhead for ``in'', ``out'' and ``in\&out'' directions is 8\%, 11\% and 21\% respectively for transferring 16 KB data, due to the extra memory copy. For OCALLs, the overhead is
negligible, since it allocates a buffer on the marshalling buffer without additional memory copy (Sec.~\ref{subsec:enclavesdk}). We remind that data transfers in ECALLs contribute a small portion to the processing time for many real-world computation workloads, especially for computation or memory intensive tasks.




\ignore{
这是因为在 ECALL 时，我们的设计需要将参数先拷贝到 marshalling buffer，再在 enclave mode 将其拷贝到安全内存，而原始 SGX SDK 是通过在 enclave mode 直接访问不安全应用的数据，将参数拷贝到安全内存，因此多一次额外的拷贝。但在 OCALL 时，我们只是将原来在不安全应用的栈上分配 buffer 改为了在 marshalling buffer 上分配，不需要额外的拷贝。In real world use-cases (see below), OCALLs are more frequency than ECALLs, we stress that the overall overhead is acceptable.
Moreover, the design brings additional protection against malicious enclave programs (Sec.~\ref{sec:defense-against-mapping-attack}).
}

\subsection{Real-world Workloads}
\label{subsec:realworld-workloads}

The evaluations were conducted on four real-world applications: an algorithm benchmark suite \textbf{NBench}~\cite{nbench}, a lightweight web server \textbf{Lighttpd}~\cite{lighttpd}, two  popular databases \textbf{SQLite}~\cite{sqlite} and \textbf{Redis}~\cite{redis}, as representations for CPU intensive, I/O-intensive, and memory-intensive tasks.
We ported the library OS Occlum~\cite{shen2020occlum} (v0.21) to the enclave SDK to reduce the porting effort
for Lighttpd and Redis.
We measured the performance on both HyperEnclave and Intel SGX, using the same code compiled under the SDK simulation mode as the baseline (providing no security guarantees). 




\ignore{
\begin{table}[t]
    \footnotesize\centering
    \begin{tabular}{lrr}
        \toprule
        Run mode   & Iterations/sec & Slowdown  \\
        \midrule
        Native      & 1902.00       & -         \\
        Simulation  & 830.35        & 56.3\%    \\
        Hardware    & 800.68        & 57.9\%    \\
        \bottomrule
    \end{tabular}
    \caption{The NBench \textsf{STRING\_SORT} test performance on Intel SGX, run in native, Simulation and Hardware mode (higher is better).}
    \label{tab:nbench}
\end{table}
}

\mypara{NBench}
NBench measures the performance of a system's CPU, FPU, and memory system, without I/O and system calls involved. We used
an adaptation of NBench to SGX, i.e., SGX-NBench~\cite{sgx-nbench} with no source code modification
for our evaluation.
As shown in Figure~\ref{fig:nbench}, the overhead introduced by HyperEnclave and SGX is about 1\% and 3\% respectively.

\mypara{SQLite} We ported SQLite (v3.19.3) with the enclave SDK, and evaluated it on both Intel SGX and HyperEnclave (GU-Enclave and HU-Enclave) using the YCSB~\cite{ycsb} workload A (50\% reads, 50\% updates).
In this evaluation we focused on the memory performance, so we configured the database as in-memory and embedded the client into the enclave to avoid I/O operations.
We increased the number of records and measured the time for 100,000 database operations.
As shown in Figure~\ref{fig:sqlite}, 
on SGX the throughput is about 75\% of the baseline for small memory usage. When the memory usage exceeds the EPC size (about 90 MB), the performance drops to 50\% due to page swapping.
On HyperEnclave, both GU-Enclave and HU-Enclave have almost the same performance as the baseline (< 5\% overhead). 
We speculate that it's because the memory encryption performance for AMD SME (without integrity protection) is faster than SGX.  

\mypara{Lighttpd}
We ran a Lighttpd (v1.4.40) server with Occlum on both SGX and HyperEnclave (GU-Enclave and HU-Enclave modes). We used the Apache HTTP benchmarking tool~\cite{ab} and ran 100 concurrent clients over the local loopback to fetch various sizes of web pages to evaluate the throughput.
In this evaluation, the overhead mainly comes from the frequent enclave mode switches (Table~\ref{tab:primitive}).
As shown in Figure~\ref{fig:lighttpd}, HU-Enclave delivers the best performance as expected (81\%  $\sim$ 88\% of the baseline). GU-Enclave achieves 69\% $\sim$ 78\% of the baseline, while SGX achieves 51\% $\sim$ 63\% of the baseline.

\mypara{Redis}
We use Redis to evaluate the performance under the comprehensive scenarios where both memory and I/O are intensive.
We ran a Redis (v6.0.9) database server with Occlum on both SGX and HyperEnclave (GU-Enclave and HU-Enclave modes).
Similar to SQLite, we configured the database as in-memory and used the YCSB workload A.
For the evaluation, we first loaded 50,000 records (in total 50 MB data) and then performed 100,000 operations from 20 clients over the local loopback.
We increased the request frequency and measured the latency under different throughput.
As shown in Figure~\ref{fig:redis}, HU-Enclave achieves 89\% of the maximum throughput of the baseline, while GU-Enclave and SGX are about 72\% and 48\% of the baseline, respectively.

\section{Discussions}
\label{sec:discuss}

\para{\atc{HyperEnclave on other platforms}}
\atc{HyperEnclave requires the virtualization extension (specifically, two-level address translation) for isolation and TPM for the root of trust and randomness, etc. Virtualization is supported on many ARM servers (such as the ARMv8 platforms~\cite{armv8}). The RISC-V H-extension specification
has evolved to v0.6.1 in 2021. Both ARM and RISC-V virtualization support two-level address translation. Certain TPM products already support ARM servers. Research has been conducted to support firmware TPM on RISC-V~\cite{boubakri2021towards}. As such, signs are promising that HyperEnclave can be adapted to run on ARM and RISC-V platforms.}

\atc{However, porting HyperEnclave to ARM and RISC-V platforms requires non-trivial engineering effort, considering that the instruction set architectures (ISAs) are totally different.} Take the ARMv8 architecture as an example. The software modules can be mapped to different exception levels (ELs): The monitor mode for \visorname can be mapped to EL2; The normal mode for the \untrustedos and untrusted part of the applications can be mapped to EL1 and EL0 respectively; The secure mode for enclaves can be mapped flexibly to EL1 or EL0. Memory isolation can be supported similarly with the support of stage 2 address translations. \atc{Furthermore}, the official Intel SGX SDK only supports x86 platforms. In particular, the transitions across enclave boundaries are handled with platform-dependent assembly code, and need to be rewritten according to the application binary interface (ABI) of the targeted platforms. We leave the \atc{further exploration of adapting} HyperEnclave to other platforms as future work.

\mypara{\atc{Attack surfaces under different enclave operation modes}}
\atc{The untrusted primary OS still runs within the VM and the attack surface from the primary OS to enclaves does not change. Running the enclaves in privileged mode or in the host, however, may expose more attack surfaces to a malicious enclave. For example, it may make the enclave malware easier to escalate to host ring-0, if the enclave runs in the host already. }
\section{Related Works}
\label{sec:related}


Most existing TEEs require specific hardware or firmware changes~\cite{mckeen2013innovative,costan2016sanctum,hunt2021confidential,kaplan2016amd,tdx2020,cca2020,hunt2021confidential}. Specifically,
{CURE~\cite{bahmani2021cure} changes the CPU core to support the enclave identifier (\textsf{eid}), and modifies the system bus to support the memory and peripheral arbiters for memory and peripheral access control according to the \textsf{eid}. Then the trusted security monitor can configure the hardware primitives to support the flexible enclaves.} In contrast, HyperEnclave supports flexible enclave modes on commodity hardwares without hardware or firmware changes.


\ignore{
, to inspect every memory access

hardware to inspect every memory access in order to prevent malicious enclaves which can modify their own page tables to access other enclave's memory.}

\atc{A line of research has been conducted to build the isolated execution environment (e.g., PAL for TrustVisor and HAP for InkTag) using virtualization extensions, including TrustVisor~\cite{McCune2010TrustVisorET}, InkTag~\cite{Hofmann2013InkTagSA}, Overshadow~\cite{chen2008overshadow}, AWS Nitro Enclaves~\cite{nitroenclave}, Microsoft Defender Credential Guard~\cite{defendercredentialguard} and Hyper-V Shielded VMs~\cite{shieldedvm}.}
TrustVisor makes the PAL page table pages read-only to prevent memory mapping attacks. The design introduces much overhead during PAL registration and switches from/to PALs. Even worse, it triggers many NPT violations in high memory pressures scenarios, due to the updates to the access and dirty bits of the PAL page tables, introducing huge overhead~\cite{Lutas2017HypervisorBM}. In Inktag, the HAP page tables are managed by the hypervisor. It allows the untrusted OS to request the hypervisor to update the HAP page tables. Both TrustVisor and InkTag are susceptible to page-table-based attacks~\cite{wang2017leaky,xu2015controlled}.
\atc{Furthermore, these designs usually contain a large code base in the TCB, including device drivers, guest IO emulation, network and block device virtualization, etc. For example,}
AWS Nitro Enclaves~\cite{nitroenclave} are constructed by the host KVM and Linux, and thus the host Linux kernel is always trusted.
\atc{We made the design choice to minimize the TCB of RustMonitor, which performs basic CPU/memory virtualization and enclave management. The primary OS kernel only needs to be trusted during the boot process and is demoted after RustMonitor starts.}

Komodo~\cite{ferraiuolo2017komodo} implements an SGX-like enclave protection model in the TrustZone environment. Keystone~\cite{lee2020keystone} supports customizable TEEs on RISC-V platforms. \atc{Komodo enclaves run in secure user mode, while Keystone enclaves run in U-mode and S-mode. In comparison, HyperEnclave supports flexible enclave mode (guest ring-3, guest ring-0/ring-3, and host ring-3). Both Komodo and HyperEnclave use page-table-based memory isolation, while Keystone uses PMP (physical memory protection) for memory isolation.}




\atc{Recently, ARM introduced the Realm Management Extension (RME) in the forthcoming ARMv9-A architecture~\cite{armrme}. The monitor (running at EL3) enforces physical memory isolation among the secure, non-secure, and Realm worlds. The trusted Realm Management Monitor (RMM), which executes at EL2 in Realm security state (R-EL2), isolates the Realms from each other through the stage 2 page tables. Similar to HyperEnclave, the RMM is much simpler than a typical hypervisor and relies on the non-secure hypervisor for device emulation etc. RME requires architectural extensions, while HyperEnclave does not. Moreover, HyperEnclave decouples its trust chain from the CPU as much as possible to construct an open and cross-platform TEE.} 

\section{Conclusion}
\label{sec:conclusion}

In this paper, we proposed HyperEnclave, an open TEE that can run on various platforms with minimum hardware requirements. It supports the process-based TEE model, and SGX programs can run on HyperEnclave with little or no code changes. Moreover, HyperEnclave supports the flexible enclave operation modes to fulfill various enclave workloads. We implemented HyperEnclave on commodity AMD servers and deployed the system internally for real-world computations. We are working on the formal verification of the implementation and plan to open source to the community.





\section*{\atc{Acknowledgments}}
\atc{We would like to thank our shepherd David Cock and the anonymous reviewers for their invaluable feedback. This work was supported in part by the National Key R\&D Program of China (Grant No. 2020YFB1805402) and the National Natural Science Foundation of China (Grant No. 61802397 and No. U19A2060).}

\bibliographystyle{plain}
\bibliography{ref}

\appendix

\section{Supplementary Materials}

\subsection{Mapping Attacks}
\label{appendix:supplement}

\begin{figure*}[htbp]
    \centering
    \begin{subfigure}[b]{0.45\textwidth}
        \centering
        \includegraphics[width=\linewidth]{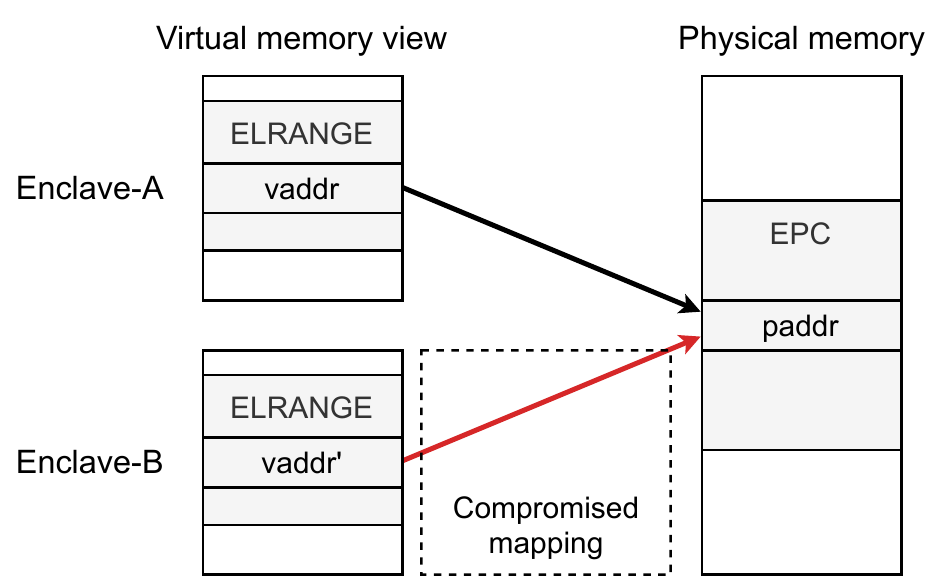}
        \caption{Alias mapping attack}
        \label{fig:alias-mapping-attack}
    \end{subfigure}
    \hspace{20pt}
    \begin{subfigure}[b]{0.45\textwidth}
        \centering
        \includegraphics[width=\linewidth]{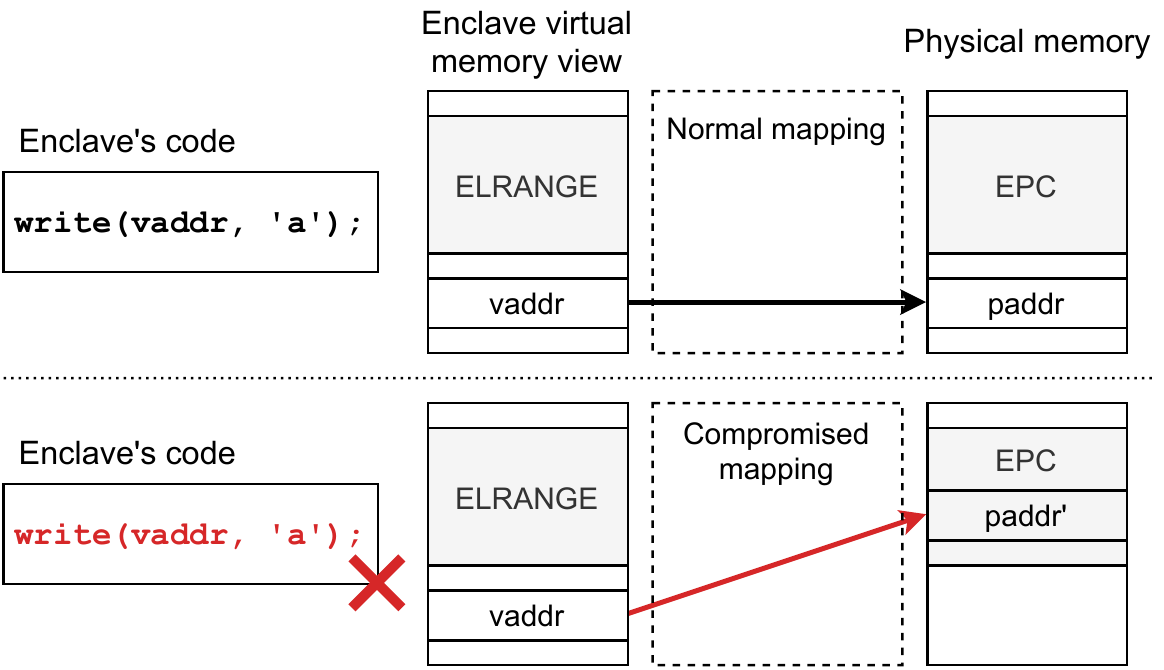}
        \caption{Remapping attack}
        \label{fig:remapping-attack}
    \end{subfigure}
    \caption{Mapping attacks. 
    (a) Two guest virtual addresses within the enclaves are mapped to the same guest physical address; (b) A non-enclave virtual address is mapped to the physical address belonging to the enclave.
    }
    \label{fig:mapping-attack}
\end{figure*}

See Figure~\ref{fig:mapping-attack}.

\subsection{Virtualization Overhead}
\label{subsec:virt-overhead}

\begin{figure}[htbp]
    \centering
    \includegraphics[width=\linewidth]{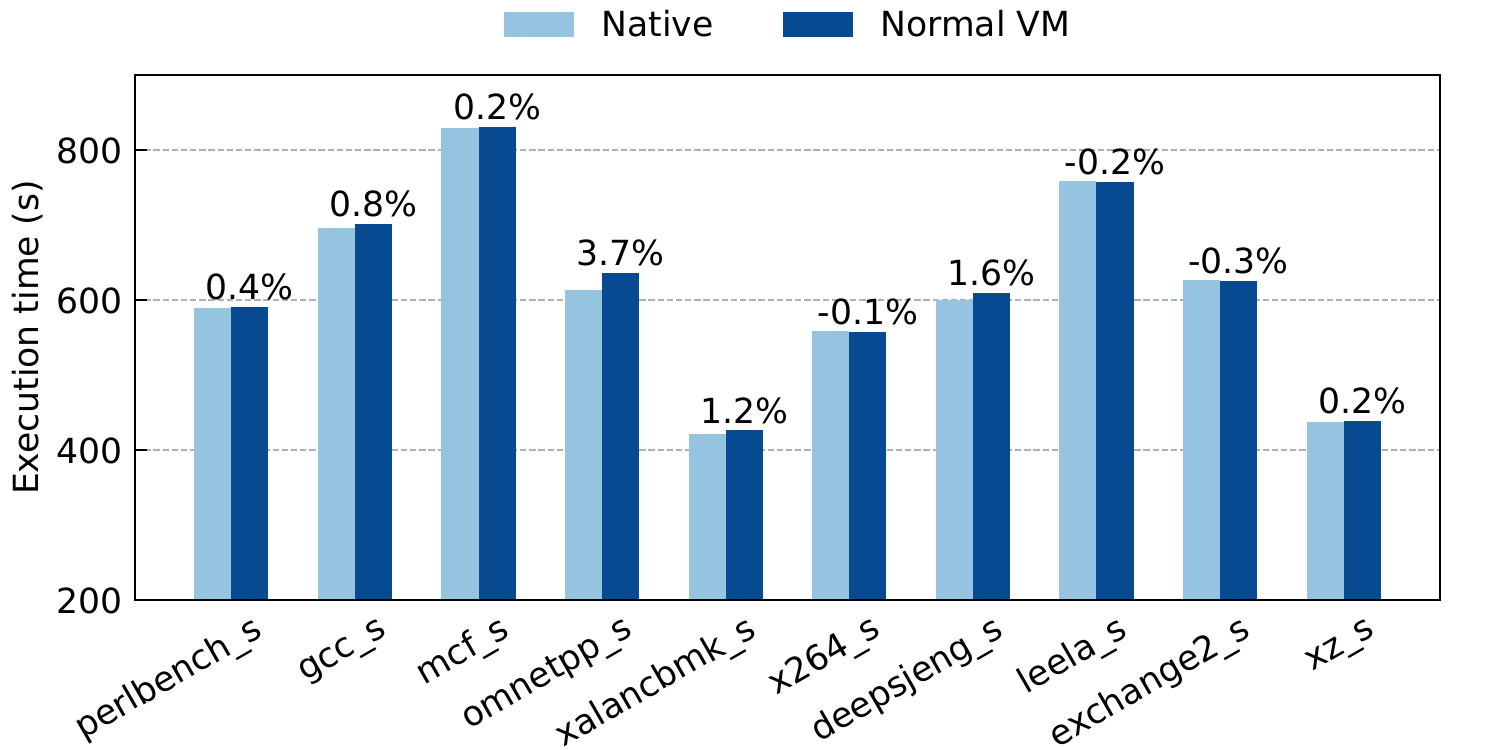}
    \caption{Virtualization overhead on SPEC CPU 2017.}
    \label{fig:speccpu}
\end{figure}

We'd like to evaluate the virtualization overhead on the normal VM. 
We ran SPEC CPU 2017 INTSpeed benchmarks~\cite{speccpu2017}, LMBench~\cite{mcvoy96lmbench}, and Linux kernel (v5.15) building when enabled and disabled \visorname.  
The result shows that the virtualization overhead is less than 1\% in most benchmarks (see Figure~\ref{fig:speccpu} and Table~\ref{tab:virt-overhead}). HyperEnclave avoids massive VM-exits by pass-through most devices to the normal VM and installs huge pages in NPT when possible to relieve the TLB pressure.

\ignore{
我们分别在启用和不启用 HyperEnclave hypervisor 的情况下，运行 SPEC CPU 2017 INTSpeed benchmarks\cite{speccpu2017}，以测量虚拟化对运行在 normal guest mode 的非 SGX 应用的影响。Figure~\ref{fig:speccpu} 给出了它们的运行时间，可见对于大部分 benchmarks 开销都小于 1\%。结果表明在启用了 HyperEnclave hypervisor 后，运行在 normal guest mode 的非 SGX 应用程序的性能几乎不受影响。这得益于 hypervisor 直通了 I/O 和中断，几乎不会发生 VM exits；且我们在将普通物理内存映射到 nested page table 时，尽可能使用了大页，大大减少了 TLB misses。
}

\begin{table}[htbp]
    \setlength{\tabcolsep}{.25em}
    \footnotesize\centering
    \begin{tabular}{r|cccccc|c}
        \toprule
                        & \multicolumn{6}{c|}{LMbench ($\mu$s)}
                        & \multirow{3}{*}{\makecell{Kernel \\ Build (s)}} \\ \cmidrule{2-7}
                        & \makecell{null \\ call}
                        & fork
                        & ctxsw
                        & mmap
                        & \makecell{Page \\ Fault}
                        & \makecell{AF \\ UNIX}
                        & \\
        \midrule
        Native          & 0.1195 & 196.3 & 3.13 & 66,125 & 0.2433 & 5.73 & 1,410 \\
        \scriptsize{Normal VM}
                        & 0.1192 & 197.9 & 3.22 & 66,407 & 0.2461 & 5.69 & 1,417 \\
        \midrule
        Overhead        & -0.25\% & 0.82\% & 2.88\% & 0.43\% & 1.15\% & -0.70\% & 0.50\% \\
        \bottomrule
    \end{tabular}
    \caption{Virtualization overhead on LMBench (null syscall, fork, context switches among 16 processes with 64KB working set, mmap, page fault, and unix socket) and building the Linux kernel.}
    \label{tab:virt-overhead}
\end{table}

\subsection{Memory Encryption Overhead}
\label{subsec:memory-overhead}

\begin{figure}[htbp]
    \centering
    \includegraphics[width=\linewidth]{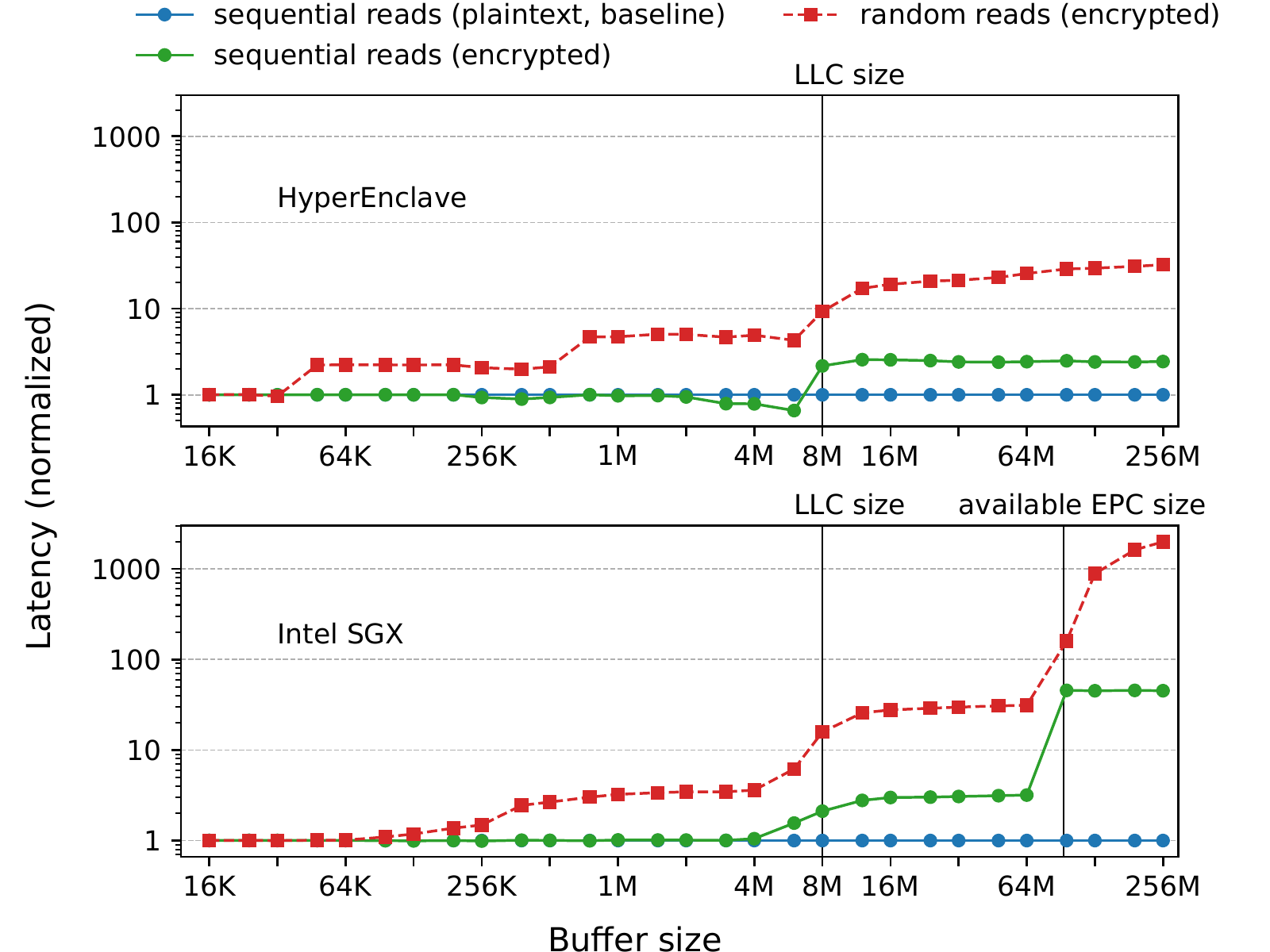}
    \caption{Memory encryption overhead for sequential and random memory accesses on HyperEnclave (with AMD SME) and Intel SGX (with Intel MEE). The LLC size is 8 MB, and the available EPC size on Intel SGX is about 93 MB.}
    \label{fig:memory-overhead}
\end{figure}

We evaluate the memory encryption overhead by measuring the memory access latency with and without encryption in  
sequential and random access patterns. The buffer size is varied from 16 KB to 256 MB.
Figure~\ref{fig:memory-overhead} illustrates the result on HyperEnclave and SGX. When the buffer size is smaller than the LLC size (8 MB), the overhead on both platforms is negligible. When the buffer size is over the LLC size, the overhead for sequential accesses and random accesses can be over 2.4$\times$ and 25$\times$ respectively on HyperEnclave, while on SGX is 3$\times$ and 30$\times$ respectively. When the buffer size exceeds the EPC size (93 MB), the overhead for sequential accesses and random accesses is 45$\times$ and 1000$\times$ slow on SGX, due to the EPC page swapping, while on HyperEnclave the overhead is still less than 30$\times$ since we reserve 24GB as enclave memory in our test. 

\section{Artifact Appendix}

\subsection*{Abstract}

HyperEnclave can support existing SGX toolchains and run SGX applications on AMD CPU with security guarantees. This artifact contains the binaries of the RustMonitor, and documentations on how to setup the HyperEnclave environment. We provide two containers to reduce the environment configuration efforts. Specifically, the server container includes the pre-installed enclave SDK, the Occlum LibOS, and the benchmarks along with their dependencies\footnote{The artifact is based on SGX SDK v2.15, Occlum LibOS v0.27, and GCC 9.4.0. The versions have little effect on the performance results.}. The client container includes pre-installed client side benchmark scripts for Lighttpd and Redis.

\subsection*{Scope}

The artifact includes benchmarks for edge calls (i.e., ECALLs and OCALLs), and benchmarks for the real-world workloads, including NBench, SQLite, Lighttpd and Redis. We provide scripts to reproduce the results in the paper (summarized in Table~\ref{tab:ae_cmp}).

\begin{table*}[t]
    \small\centering
    \begin{tabular}{C{1.6cm} C{1.6cm} C{2.1cm} C{2.1cm} C{8cm}}
        \toprule    
        Experiments   & Figure/Table    & Which container   & Estimated time &  Description     \\
        \midrule
    edge-calls & Table~\ref{tab:primitive}  & server &  10s  & \makecell[c]{The latency of \textsf{EENTER}/\textsf{EEXIT} and ECALLs/OCALLs.}  \\
    \midrule
    exception & Table~\ref{tab:exception}  & server & 20s  & Handling exceptions inside the enclaves.  \\
    \midrule
    NBench & Figure~\ref{fig:nbench}  & server & 10m  & Performance scores of NBench inside the enclaves. \\
    \midrule
    SQLite & Figure~\ref{fig:sqlite} & server & 15m & \makecell[c]{Throughput of in-memory SQLite database with different \\ number of records, under YCSB A workload.} \\
    \midrule
    Lighttpd & Figure~\ref{fig:lighttpd} & server/client &  10m    & \makecell[c]{Throughput of Lighttpd web server inside Occlum LibOS with \\ different request sizes.} \\
    \midrule
    Redis  & Figure~\ref{fig:redis}  & server/client & 20m  & \makecell[c]{Latency-throughput curve of Redis in-memory database server \\ inside Occlum LibOS with increasing request frequencies. \\ The client uses YCSB A workload.} \\
        \bottomrule
    \end{tabular}
    \caption{Summary of the benchmarks included in the artifact. }
    \label{tab:ae_cmp}
\end{table*}

\subsection*{Contents}

\begin{packeditemize}
\item{\texttt{README.md}} describes the artifact and provides a road map for evaluation.
 \item{\texttt{host/}} contains RustMonitor binary, the Linux kernel module binary, and the scripts to install and enable HyperEnclave.
 \item{\texttt{server/}} contains the source code (or patches) and scripts of all experiments to run within the enclaves. We also provide a docker container with all dependencies installed.
 \item{\texttt{client/}} contains the benchmark scripts for network-based experiments (Lighttpd and Redis) to run on the client side. We also provide a docker container with all dependencies installed.
 \item{\texttt{plots/}} contains plotting scripts to generate figures from the experiment results.
 \item{\texttt{paper-results/}} contains the results shown in the paper.
\end{packeditemize}

\subsection*{Hosting}

Check out \href{https://github.com/HyperEnclave/atc22-ae}{https://github.com/HyperEnclave/atc22-ae} (tag: \texttt{atc22-ae}, commit ID: \texttt{d1be8ab}).

\subsection*{Requirements}

\noindent\textbf{Hardware requirements}:
\begin{packeditemize}
    \item A 64-bit AMD platform with SVM enabled. Optionally, we recommend that the platform should support SME for the protection against physical memory attacks. 
    \item RAM $\ge$ 16 GB.
    \item Free disk space $\ge$ 30 GB.
\end{packeditemize}
We disabled TPM and IOMMU features in RustMonitor binary for artifact evaluation to minimize the hardware requirements. These features do not affect the performance results.

\vspace{3pt}\noindent\textbf{Software requirements}:
\begin{packeditemize}
    \item Linux with the specified kernel version (i.e., \texttt{5.3.0-28-generic}) to match our given kernel module binary. We recommend Ubuntu 18.04.4 LTS which uses this version of kernel as the default.
    \item Docker.
    \item Git.
    \item GCC and Linux kernel headers (for building the \path{enable_rdfsbase} kernel module).
\end{packeditemize}

\end{document}